\documentclass[12pt]{article}
\usepackage{geometry}
\usepackage{amsmath}
\usepackage{mathtools}
\usepackage[utf8]{inputenc}
\usepackage{float}
\usepackage{cases}
\usepackage{graphicx}
\usepackage{pgfplots}
\usepackage{pgfplotstable}
\usepackage{adjustbox}
\usepackage{amsfonts}
\usepackage{stackengine}
\usepackage{breakcites}
\usepackage{bbm}

\usepackage{stackengine}
\usepackage{caption}
\usepackage{changepage}
\usepackage{amssymb}
\usepackage{enumitem}
\usepackage{lipsum}
\DeclareMathOperator*{\argmax}{\arg\!\max}
\usepackage{tikz}
\usetikzlibrary{positioning}
\usetikzlibrary{decorations.pathreplacing,calligraphy}
\usepackage{subfig}

\usepackage[english]{babel}
\usepackage[hidelinks]{hyperref}
\hypersetup{
  colorlinks   = true, 
  urlcolor     = blue, 
  linkcolor    = blue, 
  citecolor   = blue 
}

\usepackage{ifthen}

\makeatletter
\renewcommand\@cite[2]{%
#1\ifthenelse{\boolean{@tempswa}}
{, \nolinebreak[3] #2}
}
\renewcommand\@biblabel[1]
\makeatother
\usepackage[nottoc]{tocbibind} 

\begin{document}

\title{Hierarchical Bayesian Persuasion:\\Importance of Vice Presidents\footnote{I would like to thank Ali Shourideh for his invaluable and generous support and advice.}}

\author{Majid Mahzoon\footnote{Email: \href{mailto:mmahzoon@andrew.cmu.edu}{mmahzoon@andrew.cmu.edu}}\\Tepper School of Business, Carnegie Mellon University}

\date{}

\maketitle

We study strategic information transmission in a hierarchical setting where information gets transmitted through a chain of agents up to a decision maker whose action is of importance to every agent. This situation could arise whenever an agent can communicate to the decision maker only through a chain of intermediaries, for example, an entry-level worker and the CEO in a firm, or an official in the bottom of the chain of command and the president in a government. Each agent can decide to conceal part or all the information she receives. Proving we can focus on simple equilibria, where the only player who conceals information is the first one, we provide a tractable recursive characterization of the equilibrium outcome, and show that it could be inefficient. Interestingly, in the binary-action case, regardless of the number of intermediaries, there are a few pivotal ones who determine the amount of information communicated to the decision maker. In this case, our results underscore the importance of choosing a pivotal vice president for maximizing the payoff of the CEO or president.

\section{Introduction}

Consider a hierarchical organization such as a firm or the government where the authority who takes action is the CEO or the president. Most of the times, an entry-level worker or an official in the bottom of the chain of command who has a request or suggestion, or a scandal to report, cannot communicate with the CEO/president directly. She can only communicate with her supervisor, who if desires, communicates with her own supervisor, and so on through the organization hierarchy. In other words, there is a chain of intermediaries between the entry-level worker and the CEO in the firm or between the low-ranked official and the president in the government. Knowing the communication strategy of their subordinates and the preferences of their superiors, each of these intermediaries will decide whether to pass on her information to her supervisor, and if so, to what extent. How much information will be communicated to the CEO/president? How important is the hierarchy configuration, that is, the location and preferences of each intermediary? How essential is the role of vice presidents in hierarchical organizations? 

In this paper, we are investigating the outcome of intermediated communication from a low-ranked agent to the CEO/president. It turns out regardless of the number of intermediaries, there are a few pivotal ones whose preferences and locations in the hierarchy determine the amount of information communicated to the CEO/president when facing a binary decision. In this case, our results underscore the importance of vice presidents in hierarchical organizations: By choosing an appropriate vice president, the CEO/president can maximize his payoff and avoid inefficiencies in communication.

We model this kind of hierarchical communication in the framework of Bayesian persuasion developed by \cite{7}. There is a chain of agents from an initial sender to the final receiver. The initial sender, who is the only agent with direct access to the state, intends to persuade the receiver to take some action in his favor by revealing (potentially partially) her information. However, she does not have direct access to the receiver and she can only reveal information to the next agent in the chain; the next agents in the chain do the same successively. The bottom line is that each agent, except for the first one, can only conceal all or part of the information she receives from the previous agent; she cannot reveal more information than what she receives as she does not have direct access to the state. The sequential nature of the game and the assumption that each agent, when choosing her revelation strategy, only observes the revelation strategies of the preceding players and not the information revealed by them leads us to use subgame perfect equilibrium as the solution concept.

First, we prove a version of the revelation principle for this setting: It is without loss of generality to focus on a simple class of equilibria where, on the equilibrium path, all intermediaries pass on all the information they receive. This implies that the initial sender chooses how much information she will reveal to the receiver, as in the single-sender scenario in \cite{7}, subject to that choice being incentive compatible for all intermediaries. This helps us formulate the problem of finding the most informative equilibrium outcome recursively which implies that solving a hierarchical persuasion game is equivalent to solving a single-sender persuasion game subject to recursively-defined incentive compatibility constraints. Following the restriction to simple equilibria, one may wonder if the incentive compatibility constraints can be summarized in the condition that no intermediary prefers to conceal more information. We show that while this simple condition is sufficient for incentive compatibility, it is not necessary. 

We then focus on the binary-action case where receiver will finally choose between two actions. In the special case where the state space is also binary, we fully characterize the equilibrium outcome. Existence of two agents with highly opposed preferences or an agent who would provide no information in a single-sender game would result in no information communicated to the receiver. Interestingly, if this is not the case, the relative location of two specific agents in the hierarchy determines the efficiency of the equilibrium and whether any information is communicated to the receiver in equilibrium. In this case, the degree of information revelation, if any, is determined by the preferences (bias) of one of these agents called the pivotal agent. One implication of this result is that if receiver could add an intermediary of his choice at the end of the hierarchy, i.e., the vice president, he would optimally choose her to be the new pivotal agent. The optimal choice of the vice president is (almost) independent of the hierarchy configuration and eliminates any inefficiencies in communication. 

The intuition behind the importance of vice presidents is as follows: Since agents move sequentially, every agent has to respect the preferences of the pivotal agent as long as the pivotal agent is higher in the hierarchy, i.e., closer to the receiver. By making the vice president the new pivotal player, receiver, i.e., the CEO/president, forces all agents to respect vice president's preferences.

Next, we let the state space be the interval $[0,1]$ keeping the action space binary. Focusing on simple equilibria implies that, without loss of generality, the game can be considered a binary-state, binary-action one from the intermediaries' point of view. By choosing how much information to reveal, the initial sender implicitly pins down the preferences of intermediaries in this subgame whose outcome was described above. In general, solving for the optimal choice of the initial sender is not straightforward. We fully characterize the equilibrium outcome in the special case where the prior is uniform, the differential payoff from the actions is linear in the state, and one of the actions yields zero payoff to the initial sender. 

The results in this special case are very similar to the binary-state case. Defining a binary-state game corresponding to the original game, we show that no information is communicated to the receiver in the original game if that is the case in this binary-state game. This implies that the same conditions resulting in no-information equilibrium outcome in the binary-state case are still valid in this more general case. However, if this is not the case and the prior is uniform, the relative bias of two specific agents in the hierarchy determines the efficiency of the equilibrium and whether any information is communicated to the receiver in equilibrium. In this case, every equilibrium outcome is equivalent to one which simply distinguishes between two intervals $[0,x]$ and $[x,1]$ where $x \in [0,0.5]$ is determined by the preferences (bias) of the initial sender and two agents called the pivotal agents. One implication of this result is that if receiver could add an intermediary of his choice at the end of the hierarchy, i.e., the vice president, he would optimally choose her to be one of the new pivotal agents where the optimal choice of the vice president depends on the hierarchy configuration; the new equilibrium would be efficient. Moreover, if receiver could add two players of his choice at the end of the hierarchy, he would optimally choose them to be the new pivotal agents where the order does not matter; his optimal choices are (almost) independent of the hierarchy configuration and eliminates any inefficiencies in communication. 

\vspace{2mm}

\textbf{Related Literature}. We contribute to the literature on information design and Bayesian persuasion; see \cite{3}, \cite{7}, and \cite{4}\cite{5}. We apply these tools to a setting with multiple senders who move sequentially and study how information revelation depends on senders' preferences and their order.

\cite{1} \space study the same problem under cheap talk communication while we model it in the framework of Bayesian persuasion. In fact, they extend the classic model of \cite{6} \space to investigate intermediated communication. They focus on pure-strategy equilibria and show that the set of pure strategy equilibrium outcomes does not depend on the order of intermediaries and intermediation cannot improve information in these equilibria. These results do not hold in our setting.

\cite{8}\cite{9} \space study Bayesian persuasion with multiple senders. Their main assumption is that every sender has direct access to the state and can choose a signal that is arbitrarily correlated with signals of others. They focus on pure-strategy simultaneous-move equilibria and show that greater competition, e.g., adding senders, tends to increase the amount of information revealed. In our setting where the initial sender is the only one with direct access to the state and senders move sequentially designing their experiments before observing any signal realizations, their result does not hold. In fact, \cite{11} \space show that adding senders can result in a loss of information if any of the following assumptions is violated: (i) information can be arbitrarily correlated, (ii) senders reveal information simultaneously, and (iii) senders play pure strategies.

The most closely related papers are those by \cite{12} \space and \cite{13} \space who study the sequential version of the Bayesian persuasion game with multiple senders considered by \cite{8}\cite{9}. The rich signal space considered by \cite{12} \space as well as the model considered by \cite{13} \space imply that every sender has direct access to the state and observes not only the experiments designed by the preceding senders but also the corresponding signal realizations, when designing her own experiment, and the receiver observes all designed experiments and all signal realizations; that is, each sender may only decide whether to provide more information. This is not the case in our setting where the initial sender is the only one with direct access to the state and each sender only observes the experiments designed by the preceding senders; that is, each player may only decide whether to provide less information. \cite{12} \space show that it is without loss of generality to restrict attention to a finite set of vertex beliefs. They introduce the notion of one-step equilibria which is the same as simple equilibria in our setting, and similarly formulate the incentive compatibility of intermediaries to characterize those equilibria. Investigating consultation with multiple experts, they show that adding a sender (expert) who moves first cannot reduce informativeness in equilibrium; in our setting, while this question is not relevant, the opposite holds. In fact, in our setting, which models communication in hierarchical organizations, it is more relevant to study the effect of adding a sender who moves last, i.e., the vice president; our results suggest that adding an appropriate such sender can increase informativeness and efficiency in equilibrium. \cite{13} \space uses recursive concavification to characterize the full set of subgame perfect equilibrium paths. He also introduces the notion of silent equilibria which is the same as simple equilibria in our setting; however, he does not assume a fixed tie-breaking rule for the receiver and thus provide an example of an equilibrium outcome that cannot be achieved by any silent equilibria. He also provides a sufficient condition under which it is without loss of generality to focus on silent equilibria. \cite{10} \space consider the binary-state, binary-action case of these sequential models where all senders except for the initial one care also about their reputations measured by whether the receiver's action is consistent with their recommendation. One of their results implies the importance of vice president from the initial sender's point of view, while in our setting, the vice president is important from the receiver's point of view. 

In simultaneous and independent work, \cite{2} \space also study the hierarchical Bayesian persuasion problem. They characterize the initial sender's optimal value using constrained concavification which has our notion of simple equilibria and our recursive formulation in its heart. However, the concavification method is not easy to work with; by providing a more formal characterization of recursive incentive compatibility constraints, we transform the problem into a simpler and more tractable one. This allows us to derive interesting results in the binary-action case and provide insights about the importance of vice presidents in hierarchical organizations. 


The rest of the paper is organized as follows. Section 2 sets up the model. In section 3, we prove that it is without loss of generality to focus on the set of simple equilibria where the initial sender is the only agent who may conceal information. Section 4 includes the recursive formulation of incentive compatibility constraints, restates the hierarchical persuasion problem as a single-sender one, and investigates a simple sufficient condition for the incentive compatibility conditions. In section 5, we fully characterize the equilibrium in the binary-state, binary-action case, and in a special case of the general binary-action case; we also discuss the efficiency of equilibria and importance of vice presidents. Section 6 concludes the paper. Some of the proofs are relegated to the Appendix.

\section{Model}

There are $n$ players (she) interested in the action taken by a receiver $R$ (he). Each player can try to influence the receiver's action by sending a message to the next player in a hierarchical manner. Players' and receiver's payoffs depend on the state of the world $\omega \in \Omega$ and the action taken by the receiver $a \in \mathcal{A}$ where the action space $\mathcal{A}$ is assumed to be finite. In other words, each player's (and receiver's) utility is given by a function $u_i(\omega, a)$ where $i=1,\ldots,n,R$. All players are expected utility maximizers.

The players and the receiver are uncertain about the state of the world $\omega$ and their common prior belief is represented by a probability density function $f$\footnote{If prior belief is a finite-support distribution, then we simply assume $f$ represents the probability mass function and replace all integrals with sums. The corresponding cumulative distribution function is represented by $F$.}. However, they can obtain information in a hierarchical manner as shown in Figure 1. Each player $i = 1, \ldots, n$, successively, gets to send a signal $s_i \in \mathcal{S}_i$ to player $i+1$, where $n+1 = R$; in other words, each player $i$, successively, designs (commits to) an experiment over $\mathcal{S}_{i-1}$, where $\mathcal{S}_0 = \Omega$, including a signal space $\mathcal{S}_i$ and a signal structure $\pi_i: \mathcal{S}_{i-1} \rightarrow \Delta(\mathcal{S}_i)$. In the remainder of the paper, I denote an experiment $(\mathcal{S}, \pi)$ simply by its signal structure $\pi$. Let $\Pi$ and $\Pi_i$ denote the set of all experiments over $\Omega$ and $\mathcal{S}_i$, respectively.
\begin{figure}[!h]
\centering
\begin{tikzpicture}[
conformist/.style={circle, draw=blue, fill=blue!5, very thick, minimum size=7mm}, node distance=0.5cm]

\node[conformist]   (p1)                  {\shortstack{Player 1\\ \\$\pi_1$}};
\node[conformist]   (p2)    [right=of p1] {\shortstack{Player 2\\ \\$\pi_2$}};
\node[conformist]   (p3)    [right=of p2] {\shortstack{Player 3\\ \\$\pi_3$}};
\node[conformist]   (p4)    [right=of p3] {\shortstack{Player 4\\ \\$\pi_4$}};
\node[conformist]   (p5)    [right=of p4] {\shortstack{Receiver\\ \\$a$}};

\draw[->, very thick] (p1.east) -- (p2.west);
\draw[->, very thick] (p2.east) -- (p3.west);
\draw[->, very thick] (p3.east) -- (p4.west);
\draw[->, very thick] (p4.east) -- (p5.west);

\draw[->, very thick] (-1.55,0) -- (-1.05,0);
\node[black] at (-1.3,0.3) {$\omega$};
\node[black] at (1.3,0.3) {$s_1$};
\node[black] at (3.91,0.3) {$s_2$};
\node[black] at (6.52,0.3) {$s_3$};
\node[black] at (9.13,0.3) {$s_4$};
\end{tikzpicture}
\caption{A hierarchy with $n = 4$}
\end{figure}
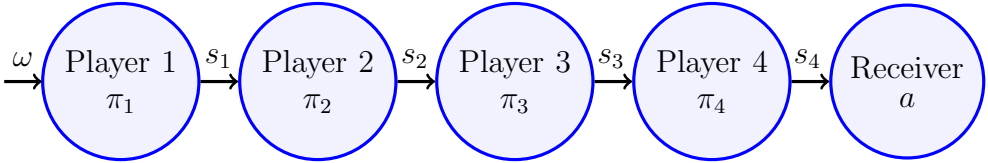

When designing her experiment, each player $i = 1, \ldots, n$ observes the experiments designed by the preceding players $\{\pi_j\}_{j=1}^{i-1}$, but not the signal realizations. After each player has designed an experiment, first the state of the world $\omega$, and then, successively, signals $s_1, \ldots, s_n$ are realized according to the designed experiments. The receiver observes the experiments designed by all players and a signal $s_n$ sent by player $n$; he updates his belief accordingly and chooses an action $a \in \mathcal{A}$ to maximize his expected utility. Note that player 1 is the only player with direct access to the state while player $n$ is the only player with direct access to the receiver.

In summary, the game consists of three stages. In the first stage, each player, successively, designs an experiment given the experiments chosen by the preceding players in order to maximize her expected utility. In the second stage, the state of the world is realized and the communication takes place in a hierarchical manner according to the designed experiments. In the last stage, the receiver updates his belief according to the designed experiments and the received signal, and decides what action to take in order to maximize his expected utility.

For the ease of exposition, we assume $\mathcal{S}_1,\mathcal{S}_2,\ldots,\mathcal{S}_n$ are all finite. As will be seen in Corollary 2, this is without loss of generality since $\mathcal{A}$ is finite. Therefore, each $\pi_i$ for $i = 1,\ldots,n$ is a finite-support conditional distribution and $\pi_i$ represents the conditional probability mass function.

\section{Equilibrium}

Given the experiments designed by all the players $(\pi_1,\ldots,\pi_n)$ and the signal $s_n$ received from player $n$, the receiver's expected utility, and thus his optimally chosen action $a^*$, only depend on his posterior belief represented by a probability density function $\mu$\footnote{If posterior belief is a finite-support distribution, then we simply assume $\mu$ represents the probability mass function.}. 

\vspace{2mm}

\textbf{Assumption 1}: If receiver is indifferent among a set of actions $A_{\mu}$ given his posterior belief $\mu$, he would take player $n$'s favorite action, i.e., $\mathbb{E}_{\mu}u_n(\omega, a^*(\mu)) \geq \mathbb{E}_{\mu}u_n(\omega, a),\ \forall a \in A_{\mu}$\footnote{Assumption 1 ensures that, in the next section, player $n$'s problem in (\ref{V_n}) is well-defined.}. 

\vspace{2mm}

Therefore, we can represent receiver's optimal strategy by $a^*: \Delta(\Omega) \rightarrow \mathcal{A}$. Given the posterior belief of the receiver $\mu$, let $v_i(\mu) = \mathbb{E}_{\mu} u_i(\omega, a^*(\mu))$ for $i=1,2,\ldots,n,R$ represent the expected utility of a player or the receiver.

The distribution of the signal observed by player $i$, that is, $s_{i-1}$, depends on the experiments designed by all the preceding players. The aggregate experiment observed by player $i$ is denoted by $\pi^{i}: \Omega \rightarrow \Delta(\mathcal{S}_{i-1})$ where 
$$\pi^i(s_{i-1}|\omega) = \sum_{s'_1} \ldots \sum_{s'_{i-2}} \pi_1(s'_1|\omega) \pi_2(s'_2|s'_1) \ldots \pi_{i-1}(s_{i-1}|s'_{i-2}).$$ 
For the ease of exposition, we write $\pi^i \equiv \pi_1 \circ \ldots \circ \pi_{i-1}$. After observing the preceding players' designed experiments, player $i$'s expected utility, and thus her optimally chosen experiment, only depend on $\pi^i$, not the individual experiments designed by the preceding players $(\pi_1,\ldots,\pi_{i-1})$. Note that by designing an experiment, each player simply garbles the aggregate experiment designed by the preceding players, that is, $\pi^{i+1} \equiv \pi^i \circ \pi_i$ is less Blackwell-informative than $\pi^i$.

Given the receiver's optimal strategy $a^*$, every profile of experiments designed by the players $(\pi_1,\ldots,\pi_n)$ induces a conditional distribution over the actions taken by the receiver $\pi: \Omega \rightarrow \Delta(\mathcal{A})$ called the outcome of the game. Without loss of generality, let's assume $\mathcal{S}_n = \mathcal{A}$ and that the receiver's optimal action is equal to the signal $s_n$ sent by player $n$\footnote{This is the case due to Assumption 1. The proof is the same as that of Proposition 1 in \cite{6}.}: $a^*(\mu_{s_n}) = s_n,\ \forall s_n \in \mathcal{S}_n$. Therefore, the outcome of the game is given by 
$$\pi(a|\omega) = \sum_{s'_1} \ldots \sum_{s'_{n-1}} \pi_1(s'_1|\omega) \pi_2(s'_2|s'_1) \ldots \pi_n(a|s'_{n-1}),\ \forall a \in \mathcal{A}, \forall \omega \in \Omega,$$ which is the aggregate experiment observed by the receiver. For the ease of exposition, we write $\pi \equiv \pi_1 \circ \ldots \circ \pi_n$\footnote{Note that $\pi = \pi^{n+1}$.}. It is now clear that the receiver's posterior belief is given by the probability density function $\mu$ where
\begin{equation*}
    \mu(\omega|\pi_1,\ldots,\pi_n,s_n) = \frac{f(\omega) \pi(s_n|\omega)}{\int_{\Omega} f(\omega') \pi(s_n|\omega') d\omega'}.
\end{equation*}

Furthermore, every outcome of the game $\pi$ induces a distribution of posteriors for the receiver $\tau_{\pi}$. We sometimes refer to $\tau_{\pi}$ as the outcome of the game. Since $\mathcal{S}_n = \mathcal{A}$ is finite, $\tau_{\pi}$ is a finite-support distribution and we simply let $\tau_{\pi}$ represent the probability mass function. Note that the above assumption implies that each signal $s_n$ sent by player $n$ induces a distinct posterior belief for the receiver $\mu_{s_n}$. Therefore,
$$\tau_{\pi}(\mu_a) = \int_{\Omega} \sum_{s'_1} \ldots \sum_{s'_{n-1}} f(\omega) \pi_1(s'_1|\omega) \pi_2(s'_2|s'_1) \ldots \pi_n(a|s'_{n-1}) d\omega = \int_{\Omega} f(\omega) \pi(a|\omega) d\omega.$$

\vspace{2mm}

\textbf{Definition 1. (Subgame Perfect Equilibria)} A subgame perfect equilibrium of the game $\sigma^* = (\pi_1^*,\sigma_2^*,\ldots,\sigma_n^*)$ is defined by an experiment $\pi_1^* \in \Pi$ for player 1 and a function (strategy) $\sigma^*_i: \Pi \rightarrow \Pi_{i-1}$ for each player $i = 2, \ldots, n$ such that 
\begin{itemize}
    \item given $\{\sigma^*_j\}_{j=2}^n$, $\pi_1^*$ maximizes $\mathbb{E}v_1(\mu)$, and
    \item for $i = 2, \ldots, n$, given $\{\sigma^*_j\}_{j=i+1}^n$, $\sigma^*_i(\pi^i)$ maximizes $\mathbb{E}v_i(\mu)$, $\forall \pi^i \in \Pi$.
\end{itemize} 

\vspace{2mm}

For a subgame perfect equilibrium of the game, let $\pi^*_i$ denote the experiment chosen by player $i$ on the equilibrium path. The corresponding equilibrium outcome and equilibrium distribution of posteriors for the receiver are denoted by $\pi^* \equiv \pi^*_1 \circ \ldots \circ \pi^*_n$ and $\tau^*$, respectively.

Now, we are ready to state the first result which is reminiscent of the revelation principle.

\vspace{2mm}

\textbf{Proposition 1. (Simple Equilibria)} Under Assumption 1, for every subgame perfect equilibrium of the game $\sigma^*$ with equilibrium outcome $\pi^*$, there exists an outcome-equivalent subgame perfect equilibrium $\sigma^{**}$ where player 1's experiment is given by $\pi_1^{**} = \pi^*$ 
and all other players use the following strategy:

$$\sigma_i^{**}(\pi^i) =
\left\{
	\begin{array}{ll}
		\mathcal{I}  & \mbox{if } \pi^i =  \pi^*\\
		\sigma_i^*(\pi^i) & \mbox{Otherwise }
	\end{array}
\right.$$
\noindent where $\mathcal{I}$ represents the full-revelation experiment, i.e., $\pi \circ I = \pi,\ \forall \pi \in \Pi$.

\vspace{2mm}

\textit{This equilibrium is simple in the sense that on the equilibrium path, all players except for the first one pass all the information they receive to the next player.}

\vspace{2mm}

\textbf{Proof}: If we prove that the strategy profile $\sigma^{**}$ is indeed an equilibrium strategy profile, then it is straightforward to see that this is outcome-equivalent to the equilibrium strategy profile $\sigma^*$.

Given the experiments chosen by the preceding players and the strategies of the succeeding players $(\pi_1, \ldots, \pi_{i-1}, \sigma_{i+1}, \ldots, \sigma_{n})$, by choosing experiment $\pi_i$, player $i$ essentially chooses the outcome of the game where her choice set is a feasible subset of $\Pi$. More explicitly, her choice set includes all outcomes of the game $\pi$ such that $\pi \equiv \pi_1 \circ \ldots \circ \pi_{i-1} \circ \pi_i \circ \sigma_{i+1}(\pi^{i+1}) \circ \ldots \circ \sigma_n(\pi^n)$ for some $\pi_i \in \Pi_{i-1}$.

\begin{itemize}

    \item Consider player $n$. The only change in her strategy compared to $\sigma^*_n$ is when she encounters a profile of experiments chosen by the preceding players such that $\pi^n = \pi^*$. In this case, she can induce any outcome $\pi$ less Blackwell-informative than $\pi^* = \pi^*_1 \circ \ldots \circ \pi^*_n$. In the equilibrium corresponding to $\sigma^*$, when she could induce any outcome $\pi$ less Blackwell-informative than $\pi^*_1 \circ \ldots \circ \pi^*_{n-1}$, including $\pi^*$, she chooses to induce $\pi^*$. Obviously, the choice set is now smaller but includes the optimal choice of a larger choice set\footnote{Blackwell information ranking is transitive.}; therefore, it is still optimal for player $n$ to induce $\pi^*$; that is, $\sigma^{**}_n(\pi^*) = \mathcal{I}$ is optimal.
    
    \item Consider player $i$ for $i=2,\ldots,n-1$. The only change in her strategy compared to $\sigma^*_i$ is when she encounters a profile of experiments chosen by the preceding players such that $\pi^i = \pi^*$. In this case, she can induce any outcome $\pi$ such that $\pi = \pi^*_1 \circ \ldots \circ \pi^*_{i-1} \circ \pi_i \circ \sigma^{**}_{i+1}(\pi^{i+1}) \circ \ldots \circ \sigma^{**}_n(\pi^n)$ by choosing $\pi_i$ where $\pi_i$ is strictly less Blackwell-informative than $\pi^*_i \circ \ldots \circ \pi^*_n$; she can also induce $\pi = \pi^*$. In the equilibrium corresponding to $\sigma^*$, when she could induce any outcome $\pi$ of the form $\pi^*_1 \circ \ldots \circ \pi^*_{i-1} \circ \pi_i \circ \sigma^*_{i+1}(\pi^{i+1}) \circ \ldots \circ \sigma^*_n(\pi^n)$ by choosing $\pi_i$, she chooses to induce $\pi^*$. Obviously, the choice set is now smaller but includes the optimal choice of a larger choice set; therefore, it is still optimal for player $i$ to induce $\pi^*$; that is, $\sigma^{**}_i(\pi^*) = \mathcal{I}$ is optimal.
    
    \item Consider player 1. By choosing $\pi_1$, she can either choose any outcome $\pi$ such that $\pi = \pi_1 \circ \sigma^{**}_2(\pi^2) \circ \ldots \circ \sigma^{**}_n(\pi^n) \neq \pi^*$, or choose $\pi = \pi^*$. In the equilibrium corresponding to $\sigma^*$, when she could induce any outcome $\pi$ of the form $\pi_1 \circ \sigma^*_2(\pi^2) \circ \ldots \circ \sigma^*_n(\pi^n)$ by choosing $\pi_1$, she chooses to induce $\pi^*$. Obviously, the choice set is the same; therefore, it is still optimal for player 1 to induce $\pi^*$; that is, $\pi^{**}_1 = \pi^*$ is optimal.
    
\end{itemize} \hfill$\blacksquare$

\vspace{2mm}

We can apply the same argument off the equilibrium path as well. Given a subgame perfect equilibrium of the game $\sigma^*$, let $\sigma^*_i(h_j)$ represent the equilibrium strategy of player $i \geq j$ following a history of the game $h_j = (\pi_1, \ldots, \pi_{j-1})$  (alternatively, in the subgame starting from player $j$). Similarly, let $\pi^*_i(h_j)$ denote the experiment chosen by player $i \geq j$ on the equilibrium path of the continuation game (alternatively, subgame).
 
\vspace{2mm} 
 
\textbf{Corollary 1.} Under Assumption 1, for every subgame perfect equilibrium of the game $\sigma^*$ with equilibrium outcome $\pi^*$, there exists an outcome-equivalent subgame perfect equilibrium $\sigma^{**}$ where following any history of the game $h_j = (\pi_1, \ldots, \pi_{j-1})$ (alternatively, in the subgame starting from player $j$), player $j$'s strategy is given by $\sigma^{**}_j(h_j) = \pi^*_j(h_j) \circ \pi^*_{j+1}(h_j) \circ \ldots \circ \pi^*_n(h_j)$ and all succeeding players use the following strategy:

$$\footnotesize \sigma_i^{**}(h_j,\pi_{j},\ldots,\pi_{i-1}) =
\left\{
	\begin{array}{ll}
		\mathcal{I}  & \mbox{if } \pi_j \circ \pi_{j+1} \circ \ldots \circ \pi_{i-1} = \\  
		& \ \ \pi^*_j(h_j) \circ \pi^*_{j+1}(h_j) \circ \ldots \circ \pi^*_n(h_j)\\
		\pi^*_i(h_i) \circ \pi^*_{i+1}(h_i) \circ \ldots \circ \pi^*_n(h_i) & \mbox{Otherwise }
	\end{array}
\right.$$

\noindent where $h_i = (h_j,\pi_{j},\ldots,\pi_{i-1})$.

\vspace{2mm}

\textit{This equilibrium is simple in the sense that following any history of the game, all players except for the first one to play pass all the information they receive to the next player.}

\vspace{2mm}

The fact that we can focus on simple equilibria implies that, on the equilibrium path, the first player recommends to the receiver which action to take, all the other players pass on that recommendation, and the receiver takes the recommended action\footnote{Similarly, following any history of the game, the first player to play recommends to the receiver which action to take and so on.}.

\textbf{Corollary 2}: Under Assumption 1, it is without loss of generality to assume $\mathcal{S}_1 = \ldots = \mathcal{S}_n = \mathcal{A}$.

\section{Incentive Compatibility Constraints and Recursive Formulation}

Given Proposition 1, in order to find the equilibrium outcome of the game $\pi^*$, or equivalently, the equilibrium distribution of posteriors for the receiver $\tau^*$, we can simply focus on the ``simple" equilibria where the first player chooses the outcome and others simply pass on the information. However, the chosen outcome must be Bayes plausible and it must be incentive compatible for other players to pass on that information. In this section, we first formally characterize these conditions, and then, formulate the problem as a set of recursive optimization problems.

Consider a pair of experiments $\pi,\pi' \in \Pi$ and their corresponding distributions of posteriors $\tau, \tau' \in \Delta(\Delta(\Omega))$. An experiment $\pi$ is more Blackwell-informative than $\pi'$ if and only if $\tau'$ is smaller than $\tau$ in the convex order: $\tau' \leq_{cx} \tau$. This means for all convex functions $\phi:\Delta(\Omega) \rightarrow \mathbb{R}$, 
$$\sum_{\mu \in supp(\tau')} \phi(\mu) \tau'(\mu) \leq \sum_{\mu \in supp(\tau)} \phi(\mu) \tau(\mu)\footnote{In the remainder of the paper, for the ease of exposition, we write $\sum_{\mu}$ instead of $\sum_{\mu \in supp(\tau)}$.}.$$
    
The following assumption simplifies the analysis by focusing on the most informative equilibrium outcome if there exist multiple Blackwell-comparable ones.\footnote{This can also be considered as focusing on the equilibrium outcome preferred by player 1 among Blackwell-comparable ones.}
    
\vspace{2mm}

\textbf{Assumption 2}: When indifferent among some experiments (revelation strategies), a player chooses the most Blackwell-informative one (reveals as much information as possible).
    
\vspace{2mm}

Now, I formally characterize the necessary conditions for the equilibrium outcome of the game:

\begin{itemize}
    \item The equilibrium outcome of the game $\tau^*$ must be Bayes plausible, i.e., $\sum_{\mu} \mu \tau^*(\mu) = f$\footnote{The proof is the same as that of Proposition 1 in \cite{6}.}. Let $\Gamma_0$ denote the set of Bayes-plausible distributions of posteriors.

    \item As mentioned before, by designing an experiment, the last player simply garbles the aggregate experiment designed by the preceding players. Therefore, in addition to the above condition, the last player should not prefer any outcome less Blackwell-informative than the equilibrium outcome $\pi^*$. Equivalently, she should not prefer any distribution of posteriors $\tau'$ that is smaller than the equilibrium distribution of posteriors $\tau^*$ in the convex order, i.e.,
    \begin{align}
    \label{IC_n}
        \tau^* \in \Gamma_0,\ \ \ \sum_{\mu} \tau'(\mu) v_n(\mu) \leq \sum_{\mu} \tau^*(\mu) v_n(\mu) ,\ \forall \tau' \leq_{cx} \tau^*.
    \end{align}
    Let $\Gamma_n \subseteq \Gamma_0$ denote the set of distributions of posteriors $\tau$ satisfying (\ref{IC_n})\footnote{Note that $\tau^* \in \Gamma_0$ and $\tau' \leq_{cx} \tau^*$ imply that $\tau' \in \Gamma_0$.}.
    
    Accordingly, we can define a value function representing the highest attainable expected utility of player $n$ as a function of the distribution of posteriors $\tau^{n}$ corresponding to the aggregate experiment $\pi^n$ designed by the preceding players:
    \begin{equation}
    \label{V_n}
        V_n(\tau^n) = \max_{\tau' \leq_{cx} \tau^n} \sum_{\mu} \tau'(\mu) v_n(\mu).
    \end{equation}
    This implies that $\Gamma_n = \{\tau \in \Gamma_0|\sum_{\mu} \tau(\mu) v_n(\mu) = V_n(\tau)\}$.
    
    \item Again, by designing an experiment, the second to last player simply garbles the aggregate experiment designed by the preceding players. Moreover, the only outcomes the second to last player can induce are those in $\Gamma_n$. Therefore, in addition to (\ref{IC_n}), the second to last player should not prefer any ``induceable" outcome less Blackwell-informative than the equilibrium outcome $\pi^*$. Equivalently, she should not prefer any distribution of posteriors $\tau' \in \Gamma_n$ that is smaller than the equilibrium distribution of posteriors $\tau^*$ in the convex order, i.e.,
    \begin{align}
    \label{IC_n-1}
        \tau^* \in \Gamma_n,\ \ \ \sum_{\mu} \tau'(\mu) v_{n-1}(\mu) \leq \sum_{\mu} \tau^*(\mu) v_{n-1}(\mu),\ \forall \tau' \in \Gamma_n\ \text{s.t.}\ \tau' \leq_{cx} \tau^*.
    \end{align}
    Let $\Gamma_{n-1} \subseteq \Gamma_n$ denote the set of distributions of posteriors $\tau$ satisfying (\ref{IC_n-1}).
    
    Accordingly, given $V_n$, we can define a value function representing the highest attainable expected utility of player $n-1$ as a function of the distribution of posteriors $\tau^{n-1}$ corresponding to the aggregate experiment $\pi^{n-1}$ designed by the preceding players:
    \begin{align}
    \label{V_n-1}
        V_{n-1}(\tau^{n-1}) = & \max_{\tau' \leq_{cx} \tau^{n-1}} \sum_{\mu} \tau'(\mu) v_{n-1}(\mu) \ \text{subject to} \\
        & \sum_{\mu} \tau'(\mu) v_n(\mu) = V_n(\tau') \nonumber
    \end{align}
    This implies that $\Gamma_{n-1} = \{\tau \in \Gamma_n|\sum_{\mu} \tau(\mu) v_{n-1}(\mu) = V_{n-1}(\tau)\}$.
    
    \item Continuing like this, now consider player $i$ for $i = 2,\ldots,n-2$. Yet Again, by designing an experiment, player $i$ simply garbles the aggregate experiment designed by the preceding players. Moreover, the only outcomes player $i$ can induce are those in $\Gamma_{i+1}$. Therefore, in addition to $\tau^* \in \Gamma_{i+1}$, player $i$ should not prefer any ``induceable" outcome less Blackwell-informative than the equilibrium outcome $\pi^*$. Equivalently, she should not prefer any distribution of posteriors $\tau' \in \Gamma_{i+1}$ that is smaller than the equilibrium distribution of posteriors $\tau^*$ in the convex order, i.e.,
    \begin{align}
    \label{IC_i}
        \tau^* \in \Gamma_{i+1},\ \ \ \sum_{\mu} \tau'(\mu) v_{i}(\mu) \leq \sum_{\mu} \tau^*(\mu) v_{i}(\mu),\ \forall \tau' \in \Gamma_{i+1}\ \text{s.t.}\ \tau' \leq_{cx} \tau^*.
    \end{align}
    Let $\Gamma_i \subseteq \Gamma_{i+1}$ denote the distributions of posteriors $\tau$ satisfying (\ref{IC_i}).
    
    Accordingly, given $V_{i+1}$, we can define a value function representing the highest attainable expected utility of player $i$ as a function of the distribution of posteriors $\tau^i$ corresponding to the aggregate experiment $\pi^i$ designed by the preceding players:
    \begin{align}
    \label{V_i}
        V_i(\tau^i) = & \max_{\tau' \leq_{cx} \tau^i} \sum_{\mu} \tau'(\mu) v_i(\mu) \ \text{subject to} \\
        & \sum_{\mu} \tau'(\mu) v_{i+1}(\mu) = V_{i+1}(\tau') \nonumber
    \end{align}
    This implies that $\Gamma_i = \{\tau \in \Gamma_{i+1}|\sum_{\mu} \tau(\mu) v_i(\mu) = V_i(\tau)\}$.
\end{itemize}

Now, I formally formulate the problem as a set of recursive optimization problems.

\vspace{2mm}

\textbf{Proposition 2. (Equilibrium Distribution of Posteriors)} Under Assumptions 1 and 2, the equilibrium distribution of posteriors for the receiver is given by
\begin{equation}
\label{Opt}
    \tau^* \in \argmax_{\tau \in \Gamma_2} \sum_{\mu} \tau(\mu) v_1(\mu). 
\end{equation}



\vspace{2mm}

\textit{Solving a hierarchical persuasion game is equivalent to solving a single-sender persuasion game subject to recursively-defined incentive compatibility constraints.}

\vspace{2mm}

\textbf{Proof}: As mentioned before, Proposition 1 allows us to simply focus on the ``simple" equilibria where the first player chooses the outcome $\tau^*$ and others simply pass on the information, as long as $\tau^*$ is Bayes plausible and it is incentive compatible for other players to pass on the information. Based on what we have discussed so far, these conditions are equivalent to $\tau^* \in \Gamma_2$. \hfill$\blacksquare$

\subsection{Can Incentive Compatibility Constraints be Simplified?}

When formulating the incentive compatibility constraints, we mentioned that the last player should not prefer any outcome
less Blackwell-informative than the equilibrium outcome, while player $i = 2,\ldots,n-1$ should not prefer any ``induceable" outcome less Blackwell-informative than the equilibrium outcome. One may wonder whether we can simplify the incentive compatibility constraints and say player $i = 2,\ldots,n$ should not prefer any outcome less Blackwell-informative than the equilibrium outcome. To that end, let $\tilde{\Gamma}$ denote the Bayes-plausible distributions of posteriors $\tau_{\pi}$ such that no player (not considering player 1) prefers any outcome less Blackwell-informative than $\pi$: 
$$\tilde{\Gamma} = \{\tau \in \Gamma_0|\sum_{\mu} \tau(\mu) v_i(\mu) \geq \sum_{\mu} \tau'(\mu) v_i(\mu),\forall \tau' \leq_{cx} \tau, \forall i=2,\ldots,n \}.$$

The next proposition illustrates that while this simple condition implies incentive compatibility and is thus a sufficient condition, it is more strict than needed and thus not a necessary condition.

\textbf{Proposition 3}: Under Assumptions 1 and 2, $\tilde{\Gamma} \subseteq \Gamma_2$, but $\tilde{\Gamma} \neq \Gamma_2$.

\textbf{Proof}: See Appendix C. \hfill$\blacksquare$






\section{Binary-Action Games}

    We now consider a common special case where the action space is binary: $\mathcal{A} = \{0,1\}$. 
    
    Based on Corollary 2, it is without loss of generality to assume $\mathcal{S}_1 = \ldots = \mathcal{S}_n = \{0,1\}$. Thus, player 1's experiment can be written as $\pi_1: \Omega \rightarrow \mathbb{R}$ where $\pi_1(\cdot)$ represents the conditional probability of sending signal $1$. Similarly, player $i$'s experiment can be written as $\pi_i: \{0,1\} \rightarrow \mathbb{R}$ for $i = 2,\ldots,n$. Following Proposition 1 and Corollary 2, every subgame perfect equilibrium of the game is outcome-equivalent to the one where the first player recommends to the receiver what action to take and all the succeeding players pass on that recommendation.
    
    Player 1 chooses an experiment $\pi_1$, or equivalently, a Bayes-plausible distribution of posteriors $\tau_1 \in \Delta(\Delta(\Omega))$ where $supp(\tau_1) = \{q_0, q_1\}$\footnote{If player 1 chooses to provide no information, $q_0 = q_1$ and we get the no-information outcome.}. From this point on, the game is a binary-state, binary-action one. In other words, given the experiment chosen by player 1, the game played by players $2,3,\ldots,n$ is a binary-state one where the state space is given by $\hat{\Omega} = supp(\tau_1)$.
    
    We now consider the binary-state, binary-action game.
    
\subsection{Binary-State Binary-Action Games}

Let $\Omega = \mathcal{A} = \{0,1\}$. Each player's and receiver's utility is represented by four numbers: $u_i^{00} = u_i(\omega = 0, a = 0)$, $u_i^{10} = u_i(\omega = 1, a = 0)$, $u_i^{01} = u_i(\omega = 0, a = 1)$, and $u_i^{11} = u_i(\omega = 1, a = 1)$ for $i = 1,\ldots,n,R$. Let $p, \mu \in \mathbb{R}$ represent the common prior belief and receiver's posterior belief that $\omega = 1$, respectively, and let $\mu_i$ represent receiver's posterior belief at which player $i$ is indifferent between the two actions: 

$$(1 - \mu_i) u_i^{00} + \mu_i u_i^{10} = (1 - \mu_i) u_i^{01} + \mu_i u_i^{11}$$
$$\mu_i = \frac{u_i^{00} - u_i^{01}}{(u_i^{00} - u_i^{01}) + (u_i^{11} - u_i^{10})}$$

All we need to know about player $i$ is her indifference posterior belief $\mu_i$\footnote{For simplicity, I assume $\mu_i \not\in \{0,p,1\}$ and $\mu_i \neq \mu_j$ if $i \neq j$.} and her preferred action at state $\omega = 1$\footnote{The argument behind this claim can be found in Appendix A.}. As a result, players can be categorized into different types as follows:

\begin{enumerate}
    \item 0-Extremists: If $\mu_i > 1$ or $\mu_i < 0$ and player $i$ prefers $a = 0$ at $\omega = 1$, she will always prefer $a = 0$.
    \item 1-Extremists: If $\mu_i > 1$ or $\mu_i < 0$ and player $i$ prefers $a = 1$ at $\omega = 1$, she will always prefer $a = 1$.
    \item Conformists: If $0 < \mu_i < 1$ and player $i$ prefers $a = 1$ at $\omega = 1$, she will prefer $a = 0$ if $\mu < \mu_i$ and $a = 1$ if $\mu > \mu_i$ as shown in Figure 2(a). 
    
    \begin{enumerate}
        \item If $0 < \mu_i < p$, player $i$ is said to be biased\footnote{A player is biased toward action $a$ if she prefers that action given the common prior belief $p$.} toward $a = 1$; the lower $\mu_i$, the higher the bias as shown in Figure 3.
        \item If $p < \mu_i < 1$, player $i$ is said to be biased toward $a = 0$; the higher $\mu_i$, the higher the bias.
    \end{enumerate}

    \item Contrarians: If $0 < \mu_i < 1$ and player $i$ prefers $a = 0$ at $\omega = 1$, she will prefer $a = 0$ if $\mu > \mu_i$ and $a = 1$ if $\mu < \mu_i$ as shown in Figure 2(b).
    
    \begin{enumerate}
        \item If $0 < \mu_i < p$, player $i$ is said to be biased toward $a = 0$; the lower $\mu_i$, the higher the bias as shown in Figure 3.
        \item If $p < \mu_i < 1$, player $i$ is said to be biased toward $a = 1$; the higher $\mu_i$, the higher the bias.
    \end{enumerate}
    
\end{enumerate}

\begin{figure}[!h]
\centering
\subfloat[\small{Conformist}]{
    \begin{tikzpicture}
    \draw[blue, very thick] (-4,-5) -- (-1,-5);
    \draw[red, very thick] (-1,-5) -- (4,-5);
    
    \filldraw[black] (-4,-5) circle (2pt);
    \coordinate [label=below:$0$] (A) at (-4,-5.2);
    
    \filldraw[black] (4,-5) circle (2pt);
    \coordinate [label=below:$1$] (B) at (4,-5.2);
    
    \filldraw[black] (-1,-5) circle (2pt);
    \coordinate [label=below:$\mu_i$] (C) at (-1,-5.2);
    
    \draw [decorate, decoration = {brace}, thin] (-4,-4.7) --  (-1.1,-4.7);
    \node[blue] at (-2.5,-4.2) {prefers $a = 0$};
    
    \draw [decorate, decoration = {brace}, thin] (-0.9,-4.7) --  (4,-4.7);
    \node[red] at (1.5,-4.2) {prefers $a = 1$};
    \end{tikzpicture}
}

\subfloat[\small{Contrarian}]{
    \begin{tikzpicture}
    \draw[red, very thick] (-4,-5) -- (-1,-5);
    \draw[blue, very thick] (-1,-5) -- (4,-5);
    
    \filldraw[black] (-4,-5) circle (2pt);
    \coordinate [label=below:$0$] (A) at (-4,-5.2);
    
    \filldraw[black] (4,-5) circle (2pt);
    \coordinate [label=below:$1$] (B) at (4,-5.2);
    
    \filldraw[black] (-1,-5) circle (2pt);
    \coordinate [label=below:$\mu_i$] (C) at (-1,-5.2);
    
    \draw [decorate, decoration = {brace}, thin] (-4,-4.7) --  (-1.1,-4.7);
    \node[red] at (-2.5,-4.2) {prefers $a = 1$};
    
    \draw [decorate, decoration = {brace}, thin] (-0.9,-4.7) --  (4,-4.7);
    \node[blue] at (1.5,-4.2) {prefers $a = 0$};
    \end{tikzpicture}
}
\caption{Action preferences of conformists and contrarians}
\end{figure}
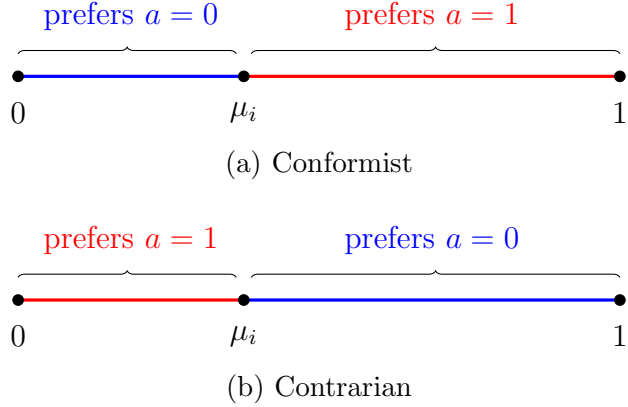

For example, consider the following utility function
\begin{equation*}
    u_i(\omega, a) = - (a - \omega) ^2 - \alpha_i a,
\end{equation*}
where $\alpha_i$ represents player $i$'s willingness (bias) to take action $a = 0$. Then we have $\mu_i = \frac{1+\alpha_i}{2}$. In this example, if $\alpha_i > 1$, player $i$ is a 0-extremist, and if $\alpha_i < -1$, she is a 1-extremist; otherwise, she is a conformist as she would prefer to match the state.

Similarly, consider the following utility function
\begin{equation*}
    u_i(\omega, a) = (a - \omega) ^2 - \alpha_i (1 - a),
\end{equation*}
where $\alpha_i$ represents player $i$'s willingness (bias) to take action $a = 1$. Then we have $\mu_i = \frac{1+\alpha_i}{2}$. In this example, if $\alpha_i > 1$, player $i$ is a 1-extremist, and if $\alpha_i < -1$, she is a 0-extremist; otherwise, she is a contrarian as she would prefer not to match the state.

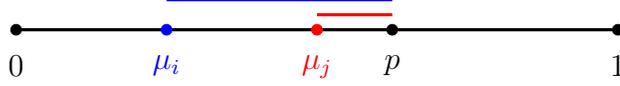
\begin{figure}[!h]
\centering
\begin{tikzpicture}
    \draw[black, very thick] (-4,-5) -- (4,-5);
    \draw[blue, very thick] (-2,-4.6) -- (1,-4.6);
    \draw[red, very thick] (0,-4.8) -- (1,-4.8);
    
    \filldraw[black] (-4,-5) circle (2pt);
    \coordinate [label=below:$0$] (A) at (-4,-5.2);
    
    \filldraw[black] (4,-5) circle (2pt);
    \coordinate [label=below:$1$] (B) at (4,-5.2);
    
    \filldraw[blue] (-2,-5) circle (2pt);
    \coordinate [label={[blue]below:$\mu_i$}] (C) at (-2,-5.2);
    
    \filldraw[red] (0,-5) circle (2pt);
    \coordinate [label={[red]below:$\mu_j$}] (D) at (0,-5.2);
    
    \filldraw[black] (1,-5) circle (2pt);
    \coordinate [label=below:$p$] (E) at (1,-5.2);
    
    
\end{tikzpicture}
\caption{The lengths of the blue and red segments represent the biases of players $i$ and $j$, respectively. Player $i$ is higher biased than player $j$.}
\end{figure}





If receiver is a 0-extremist or a 1-extremist, he will always take action $a = 0$ or $a = 1$, respectively, regardless of the experiments chosen by the players, and thus all players are indifferent among all experiments. Given Assumption 2, receiver gets full information in equilibrium.

Suppose receiver is a conformist who is biased toward $a=1$, i.e., $0 < \mu_R < p$. Define the following sets of players:
\begin{itemize}
    \item $A = \{i: i\ \text{is a conformist with}\ 0 < \mu_i < \mu_R\}$
    \item $B = \{i: i\ \text{is a conformist with}\ p < \mu_i < 1\}$
    \item $C = \{i: i\ \text{is a contrarian with}\ \mu_R < \mu_i < 1\}$
    \item $D = \{i: i\ \text{is a contrarian with}\ 0 < \mu_i < \mu_R\}$
    \item $E_0 = \{i: i\ \text{is a 0-extremist}\}$
    \item $E_1 = \{i: i\ \text{is a 1-extremist}\}$
\end{itemize}

The next proposition implies that the relative location of two specific players is relevant in determining the equilibrium outcome of the binary-state, binary-action game:
\begin{itemize}
    \item Let $A^*$ represent the player with the highest bias among those in $A$: $\mu_{A^*} = \min_{i \in A} \mu_i$. If $A = \emptyset$, let $A^* = R$. Figure 4 illustrates an example.
    
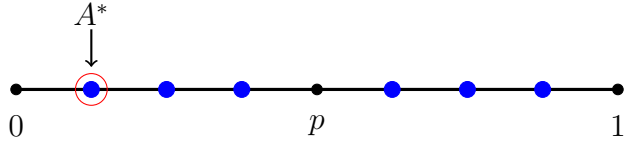
\begin{figure}[!h]
\centering
\begin{tikzpicture}
    \draw[black, very thick] (-4,-5) -- (4,-5);
    
    \filldraw[black] (-4,-5) circle (2pt);
    \coordinate [label=below:$0$] (A) at (-4,-5.2);
    
    \filldraw[black] (4,-5) circle (2pt);
    \coordinate [label=below:$1$] (B) at (4,-5.2);
    
    \filldraw[black] (0,-5) circle (2pt);
    \coordinate [label=below:$p$] (C) at (0,-5.2);
    
    \filldraw[blue] (3,-5) circle (3pt);
    \filldraw[blue] (2,-5) circle (3pt);
    \filldraw[blue] (1,-5) circle (3pt);
    \filldraw[blue] (-1,-5) circle (3pt);
    \filldraw[blue] (-2,-5) circle (3pt);
    \filldraw[blue] (-3,-5) circle (3pt);
    \draw[red] (-3,-5) circle (6pt);
    
    \draw[->, thick] (-3,-4.2) -- (-3,-4.7);
    \node[black] at (-3,-4) {$A^*$};
\end{tikzpicture}
\caption{The blue dots represent the indifference beliefs of all the conformists in the hierarchy including the receiver. The ones to the right of $p$ correspond to the conformists biased toward $a = 0$ and those to the left of $p$ correspond to the conformists biased toward $a = 1$.}
\end{figure}
    
    \item If $D \cup E_0 \neq \emptyset$, let $E^*$ represent the player closest to the receiver among those in $D \cup E_0$: $E^* = \max_{i \in D \cup E_0} i$. Figure 5 illustrates an example.
    
\begin{figure}[!h]
\centering
\begin{tikzpicture}[
conformist/.style={circle, draw=blue, fill=blue!5, very thick, minimum size=7mm},
contrarian/.style={circle, draw=red , fill=red!5 , very thick, minimum size=7mm}, node distance=0.5cm]

\node[conformist]   (p1)                  {1};
\node[contrarian]   (p2)    [right=of p1] {2};
\node[conformist]   (p3)    [right=of p2] {3};
\node[conformist]   (p4)    [right=of p3] {4};
\node[contrarian]   (p5)    [right=of p4] {5};
\node[contrarian]   (p6)    [right=of p5] {6};
\node[conformist]   (p7)    [right=of p6] {7};
\node[conformist]   (p8)    [right=of p7] {R};

\draw[->, very thick] (p1.east) -- (p2.west);
\draw[->, very thick] (p2.east) -- (p3.west);
\draw[->, very thick] (p3.east) -- (p4.west);
\draw[->, very thick] (p4.east) -- (p5.west);
\draw[->, very thick] (p5.east) -- (p6.west);
\draw[->, very thick] (p6.east) -- (p7.west);
\draw[->, very thick] (p7.east) -- (p8.west);

\draw[->, thick] (6.35,1) -- (6.35,0.5);
\node[black] at (6.35,1.2) {$E^*$};
\end{tikzpicture}
\caption{The red circles represent players in $D \cup E_0$ of a hierarchy with $n = 7$.}
\end{figure}
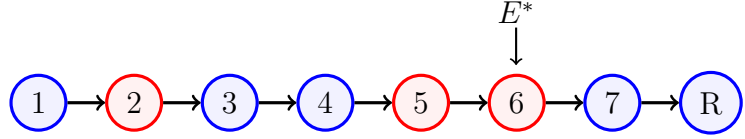

\end{itemize}

\vspace{2mm}

\textbf{Proposition 4}: Let $\Omega = \mathcal{A} = \{0,1\}$ and suppose receiver is a conformist who is biased toward $a=1$\footnote{$0 < \mu_R < p$.} and Assumption 1 holds\footnote{Assumption 2 is not necessary in the binary-state, binary-action games.}. If there exists a 1-extremist or a contrarian with $\mu_i > \mu_{A^*}$\footnote{$C \cup E_1 \neq \emptyset$ or $\exists i \in D$ such that $\mu_i > \mu_{A^*}$.}, the equilibrium is characterized by \textit{no-information} outcome\footnote{No information outcome is characterized by $supp(\tau^*) \subset (\mu_R,1]$ if $a^*(\mu_R) = 0$ or $supp(\tau^*) \subset [\mu_R,1]$ if $a^*(\mu_R) = 1$, which is determined by Assumption 1. However, under the assumption that $\mathcal{S}_n = \mathcal{A}$, no-information outcome is equivalent to $supp(\tau^*) = {p}$.}. Otherwise, 
\begin{enumerate}[label=\alph*.]
    \item if all players are conformists\footnote{$D \cup E_0 = \emptyset$.}, the equilibrium outcome $\tau^*$ is characterized by $supp(\tau^*) = \{0,1\}$ (\textit{full-information outcome}).
    \item if there are players who are not conformists and $A^* < E^*$, the equilibrium is characterized by \textit{no-information} outcome.
    \item if there are players who are not conformists and $A^* > E^*$, the equilibrium outcome $\tau^*$ is characterized by $supp(\tau^*) = \{\mu_{A^*},1\}$\footnote{Note that Assumption 1 is needed here if $A^* = R$; otherwise, there would be no equilibrium.}.
\end{enumerate}

\vspace{2mm}

\textit{In the equilibrium, if all players are conformists, receiver gets full information. Otherwise\footnote{This is the case except for some rather uninteresting cases resulting in no information for the receiver, as explained in what follows.}, location of the pivotal player $A^*$ determines whether receiver gets any information, and if so, her bias $\mu_{A^*}$ determines the amount of information communicated to the receiver as shown in Figure 6; the higher the bias of $A^*$, the more information communicated.}

\vspace{2mm}

\textbf{Proof}: See Appendix A. \hfill$\blacksquare$


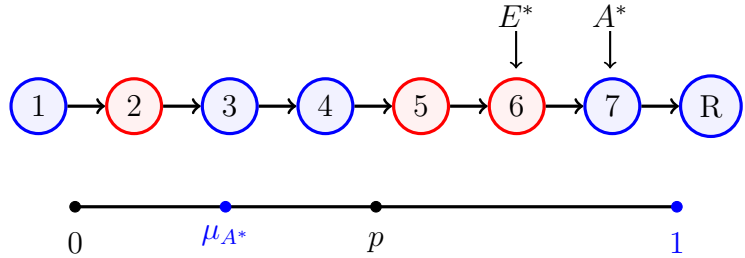
\begin{figure}[!h]
\centering
\begin{tikzpicture}[
conformist/.style={circle, draw=blue, fill=blue!5, very thick, minimum size=7mm},
contrarian/.style={circle, draw=red , fill=red!5 , very thick, minimum size=7mm}, node distance=0.5cm]

\node[conformist]   (p1)                  {1};
\node[contrarian]   (p2)    [right=of p1] {2};
\node[conformist]   (p3)    [right=of p2] {3};
\node[conformist]   (p4)    [right=of p3] {4};
\node[contrarian]   (p5)    [right=of p4] {5};
\node[contrarian]   (p6)    [right=of p5] {6};
\node[conformist]   (p7)    [right=of p6] {7};
\node[conformist]   (p8)    [right=of p7] {R};

\draw[->, very thick] (p1.east) -- (p2.west);
\draw[->, very thick] (p2.east) -- (p3.west);
\draw[->, very thick] (p3.east) -- (p4.west);
\draw[->, very thick] (p4.east) -- (p5.west);
\draw[->, very thick] (p5.east) -- (p6.west);
\draw[->, very thick] (p6.east) -- (p7.west);
\draw[->, very thick] (p7.east) -- (p8.west);

\draw[->, thick] (7.6,1) -- (7.6,0.5);
\node[black] at (7.6,1.2) {$A^*$};

\draw[->, thick] (6.35,1) -- (6.35,0.5);
\node[black] at (6.35,1.2) {$E^*$};
\end{tikzpicture}

\vspace{8mm}

\begin{tikzpicture}
    \draw[black, very thick] (-4,-5) -- (4,-5);
    
    \filldraw[black] (-4,-5) circle (2pt);
    \coordinate [label=below:$0$] (A) at (-4,-5.2);
    
    \filldraw[blue] (4,-5) circle (2pt);
    \coordinate [label={[blue]below:$1$}] (B) at (4,-5.2);
    
    \filldraw[black] (0,-5) circle (2pt);
    \coordinate [label=below:$p$] (C) at (0,-5.2);
    
    \filldraw[blue] (-2,-5) circle (2pt);
    \coordinate [label={[blue]below:$\mu_{A^*}$}] (D) at (-2,-5.1);

\end{tikzpicture}
\caption{A hierarchy with $n = 7$ and its equilibrium distribution of posteriors: The blue dots represent the support of the equilibrium distribution of posteriors.}
\end{figure}

Another specific player will be of importance in the next results: Let $D^*$ represent the player with the lowest bias among those in $D$ with $\mu_i < \mu_{A^*}$: $\mu_{D^*} = \max_{i \in D: \mu_i < \mu_{A^*}} \mu_i$. If $\{i \in D: \mu_i < \mu_{A^*}\} = \emptyset$, let $\mu_{D^*} = 0$.

\vspace{2mm}

No-information equilibrium outcome could emerge because of three reasons:
\begin{enumerate}
    \item Existence of players who would provide no information in a single-sender game: This is the case if there exists a 1-extremist or a contrarian with $\mu_R < \mu_i$ as shown in Figure 7, i.e., $C \cup E_1 \neq \emptyset$.
    
\begin{figure}[!h]
\centering
\begin{tikzpicture}
    \draw[red, very thick] (-4,-5) -- (0,-5);
    \draw[blue, very thick] (0,-5) -- (4,-5);
    
    \filldraw[black] (-4,-5) circle (2pt);
    \coordinate [label=below:$0$] (A) at (-4,-5.2);
    
    \filldraw[black] (4,-5) circle (2pt);
    \coordinate [label=below:$1$] (B) at (4,-5.2);
    
    \filldraw[black] (0,-5) circle (2pt);
    \coordinate [label=below:$\mu_i$] (C) at (0,-5.2);
    
    \filldraw[black] (-1,-5) circle (2pt);
    \coordinate [label=below:$\mu_R$] (D) at (-1,-5.2);
    
    \draw [decorate, decoration = {brace}, thin] (-4,-4.7) --  (-0.1,-4.7);
    \node[red] at (-2,-4.2) {$i$ prefers $a = 1$};
    
    \draw [decorate, decoration = {brace}, thin] (0.1,-4.7) --  (4,-4.7);
    \node[blue] at (2,-4.2) {$i$ prefers $a = 0$};

    \draw[blue, very thick] (-4,-6) -- (-1,-6);
    \draw[red, very thick] (-1,-6) -- (4,-6);
    
    \filldraw[black] (-4,-6) circle (2pt);
    \filldraw[black] (4,-6) circle (2pt);
    \filldraw[black] (0,-6) circle (2pt);
    \filldraw[black] (-1,-6) circle (2pt);
    
    \draw [decorate, decoration = {brace}, thin] (-1.1,-6.3) -- (-4,-6.3);
    \node[blue] at (-2.5,-6.8) {$a^*(\mu) = 0$};
    
    \draw [decorate, decoration = {brace}, thin]  (4,-6.3) -- (-0.9,-6.3);
    \node[red] at (1.5,-6.8) {$a^*(\mu) = 1$};
\end{tikzpicture}
\caption{The blue segments do not overlap, that is, there does not exist a posterior belief at which both player $i$ and the receiver prefer $a = 0$.}
\end{figure}
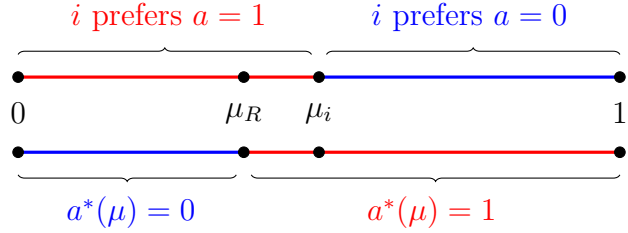
    
    \item Existence of players with highly opposed preferences: Assuming $C \cup E_1 = \emptyset$, this is the case if there exists $i \in D$ such that $\mu_i > \mu_{A^*}$ as shown in Figure 8.
    
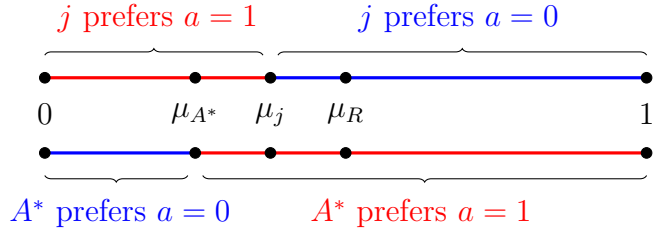
\begin{figure}[!h]
\centering
\begin{tikzpicture}
    \draw[red, very thick] (-4,-5) -- (-1,-5);
    \draw[blue, very thick] (-1,-5) -- (4,-5);
    
    \filldraw[black] (-4,-5) circle (2pt);
    \coordinate [label=below:$0$] (A) at (-4,-5.2);
    
    \filldraw[black] (4,-5) circle (2pt);
    \coordinate [label=below:$1$] (B) at (4,-5.2);
    
    \filldraw[black] (0,-5) circle (2pt);
    \coordinate [label=below:$\mu_R$] (C) at (0,-5.2);
    
    \filldraw[black] (-1,-5) circle (2pt);
    \coordinate [label=below:$\mu_j$] (D) at (-1,-5.2);
    
    \filldraw[black] (-2,-5) circle (2pt);
    \coordinate [label=below:$\mu_{A^*}$] (D) at (-2,-5.2);
    
    \draw [decorate, decoration = {brace}, thin] (-4,-4.7) --  (-1.1,-4.7);
    \node[red] at (-2.5,-4.2) {$j$ prefers $a = 1$};
    
    \draw [decorate, decoration = {brace}, thin] (-0.9,-4.7) --  (4,-4.7);
    \node[blue] at (1.5,-4.2) {$j$ prefers $a = 0$};
    
    \draw[blue, very thick] (-4,-6) -- (-2,-6);
    \draw[red, very thick] (-2,-6) -- (4,-6);
    
    \filldraw[black] (-4,-6) circle (2pt);
    \filldraw[black] (4,-6) circle (2pt);
    \filldraw[black] (0,-6) circle (2pt);
    \filldraw[black] (-1,-6) circle (2pt);
    \filldraw[black] (-2,-6) circle (2pt);
    
    \draw [decorate, decoration = {brace}, thin] (-2.1,-6.3) -- (-4,-6.3);
    \node[blue] at (-3,-6.8) {$A^*$ prefers $a = 0$};
    
    \draw [decorate, decoration = {brace}, thin]  (4,-6.3) -- (-1.9,-6.3);
    \node[red] at (1,-6.8) {$A^*$ prefers $a = 1$};
\end{tikzpicture}
\caption{The blue segments do not overlap, that is, there does not exist a posterior belief at which both player $i$ and $A^*$ prefer $a = 0$.}
\end{figure}

    \item Location of the pivotal player: Assuming $C \cup E_1 = \emptyset$ and there exists no $i \in D$ such that $\mu_i > \mu_{A^*}$, this is the case if $A^* < E^*$, i.e., $E^*$ is closer to the receiver than the pivotal player $A^*$ as shown in Figure 9. This equilibrium is inefficient since all players prefer every outcome $\tau$ with $\min(supp(\tau)) \in [\mu_{D^*},\mu_{A^*}]$. The equilibrium is inefficient if and only if this condition holds.
    
\begin{figure}[!h]
\centering
\begin{tikzpicture}[
conformist/.style={circle, draw=blue, fill=blue!5, very thick, minimum size=7mm},
contrarian/.style={circle, draw=red , fill=red!5 , very thick, minimum size=7mm}, node distance=0.5cm]

\node[conformist]   (p1)                  {1};
\node[contrarian]   (p2)    [right=of p1] {2};
\node[conformist]   (p3)    [right=of p2] {3};
\node[conformist]   (p4)    [right=of p3] {4};
\node[contrarian]   (p5)    [right=of p4] {5};
\node[contrarian]   (p6)    [right=of p5] {6};
\node[conformist]   (p7)    [right=of p6] {7};
\node[conformist]   (p8)    [right=of p7] {R};

\draw[->, very thick] (p1.east) -- (p2.west);
\draw[->, very thick] (p2.east) -- (p3.west);
\draw[->, very thick] (p3.east) -- (p4.west);
\draw[->, very thick] (p4.east) -- (p5.west);
\draw[->, very thick] (p5.east) -- (p6.west);
\draw[->, very thick] (p6.east) -- (p7.west);
\draw[->, very thick] (p7.east) -- (p8.west);

\draw[->, thick] (3.8,1) -- (3.8,0.5);
\node[black] at (3.8,1.2) {$A^*$};

\draw[->, thick] (6.35,1) -- (6.35,0.5);
\node[black] at (6.35,1.2) {$E^*$};
\end{tikzpicture}
\caption{The blue circles represent the conformists while the red ones represent players in $\{i \in D: \mu_i < \mu_{A^*}\} \cup E_0$ of a hierarchy with $n = 7$.}
\end{figure}
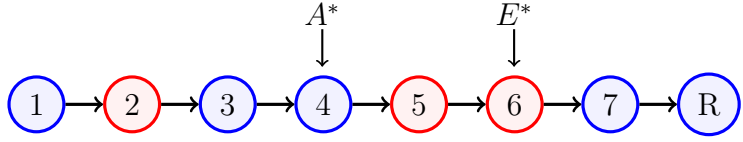

\end{enumerate}

The next corollary shows that the configuration 3 above is the only one which causes inefficiency in communication.

\vspace{2mm}

\textbf{Corollary 3}: Let $\Omega = \mathcal{A} = \{0,1\}$ and suppose receiver is a conformist who is biased toward $a=1$ and Assumption 1 holds. The equilibrium is inefficient if and only if the conditions of Proposition 4.b. hold.

\vspace{2mm}

The reason is that when designing her experiment, $E^*$, whose preferences are somewhat opposed to those of $A^*$, only considers the incentive compatibility constraints of the succeeding players, not those of the preceding players such as $A^*$. Knowing this, $A^*$ preemptively provides no information.

One question that may arise is if the receiver, who could be the CEO of a firm or the president, can get more information and thus increase her payoff by assigning a vice president and if so, how. The next corollary shows that the answer is affirmative.

\vspace{2mm}

\textbf{Corollary 4}: Let $\Omega = \mathcal{A} = \{0,1\}$ and suppose receiver is a conformist who is biased toward $a=1$ and Assumption 1 holds. If receiver could add a player of his choice at the end of the hierarchy, he would choose a conformist biased toward $a=1$ with higher bias than $A^*$ but lower bias than $D^*$\footnote{$\mu_{D^*} < \mu_i < \mu_{A^*}$.}. Receiver's utility would then be increasing in the bias of the added player and the new equilibrium would be efficient. Moreover, he does not benefit from adding more players.

\vspace{2mm}

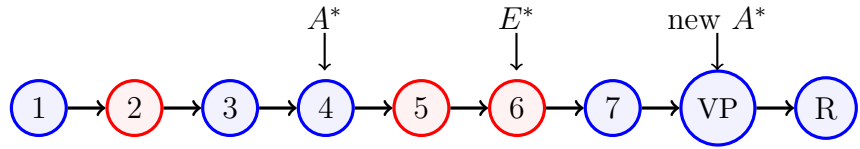
\begin{figure}[!h]
\centering
\begin{tikzpicture}[
conformist/.style={circle, draw=blue, fill=blue!5, very thick, minimum size=7mm},
contrarian/.style={circle, draw=red , fill=red!5 , very thick, minimum size=7mm}, node distance=0.5cm]

\node[conformist]   (p1)                  {1};
\node[contrarian]   (p2)    [right=of p1] {2};
\node[conformist]   (p3)    [right=of p2] {3};
\node[conformist]   (p4)    [right=of p3] {4};
\node[contrarian]   (p5)    [right=of p4] {5};
\node[contrarian]   (p6)    [right=of p5] {6};
\node[conformist]   (p7)    [right=of p6] {7};
\node[conformist]   (new)   [right=of p7] {\small{VP}};
\node[conformist]   (p8)    [right=of new] {R};

\draw[->, very thick] (p1.east) -- (p2.west);
\draw[->, very thick] (p2.east) -- (p3.west);
\draw[->, very thick] (p3.east) -- (p4.west);
\draw[->, very thick] (p4.east) -- (p5.west);
\draw[->, very thick] (p5.east) -- (p6.west);
\draw[->, very thick] (p6.east) -- (p7.west);
\draw[->, very thick] (p7.east) -- (new.west);
\draw[->, very thick] (new.east) -- (p8.west);

\draw[->, thick] (3.8,1) -- (3.8,0.5);
\node[black] at (3.8,1.2) {$A^*$};

\draw[->, thick] (6.35,1) -- (6.35,0.5);
\node[black] at (6.35,1.2) {$E^*$};

\draw[->, thick] (9.02,1) -- (9.02,0.5);
\node[black] at (9.02,1.2) {new $A^*$};

\end{tikzpicture}
\caption{Adding a vice president to a hierarchy with $n = 7$}
\end{figure}

Essentially, every player is better off by respecting the preferences of the pivotal player as long as the pivotal player is closer to the receiver. By behaving as prescribed in the corollary, receiver makes vice president the new pivotal player as shown in Figure 10, and all players are better off by respecting her preferences. As mentioned before, the higher the bias of the pivotal player, the more information communicated.

To generalize Proposition 4 to other types of the receiver, we call conformist the opposite type of contrarian and vice versa. Similarly, we call bias toward $a=1$ the opposite of bias toward $a=0$ and vice versa. Also, when two players are of opposite type and bias, we compare their biases in the same way as if they are of the same type and bias.

Define the following sets of players:
\begin{itemize}
    \item $A = \{i: i\ \text{is of the same type and bias as receiver but with higher bias}\}$
    \item $B = \{i: i\ \text{is of the same type as receiver but with opposite bias}\}$
    \item $C = \{i: i\ \text{is of opposite type to receiver but with the same bias or lower opposite bias}\}$
    \item $D = \{i: i\ \text{is of opposite type to receiver but with higher opposite bias}\}$
    \item $E_0 = \{i: i\ \text{is an extremist of opposite bias to receiver}\}$
    \item $E_1 = \{i: i\ \text{is an extremist of the same bias as receiver}\}$

\end{itemize}

As in Proposition 4, the relative location of two specific players is relevant in determining the equilibrium outcome of the binary-state, binary-action game:
\begin{itemize}
    \item Let $A^*$ represent the player with the highest bias among those in $A$; if $A = \emptyset$, let $A^* = R$.
    \item If $D \cup E_0 \neq \emptyset$, let $E^*$ represent the player closest to the receiver among those in $D \cup E_0$: $E^* = \max_{i \in D \cup E_0} i$.
\end{itemize}

\vspace{2mm}

\textbf{Proposition 5}: Let $\Omega = \mathcal{A} = \{0,1\}$ and suppose receiver is not an extremist and Assumption 1 holds. If $C \cup E_1 \neq \emptyset$ or if there exists $i \in D$ such that $i$ is less biased than $A^*$, the equilibrium is characterized by \textit{no-information} outcome\footnote{No information outcome is characterized by (i) $supp(\tau^*) \subset (\mu_R,1]$ or $supp(\tau^*) \subset [\mu_R,1]$, or (ii) $supp(\tau^*) \subset [0,\mu_R)$ or $supp(\tau^*) \subset [0,\mu_R]$, whichever includes $p$; open or closed interval is determined by Assumption 1. However, under the assumption that $\mathcal{S}_n = \mathcal{A}$, no-information outcome is equivalent to $supp(\tau^*) = {p}$.}. Otherwise, 
\begin{enumerate}[label=\alph*.]
    \item if $D \cup E_0 = \emptyset$\footnote{All players are of the same type as the receiver.}, the equilibrium outcome $\tau^*$ is characterized by $supp(\tau^*) = \{0,1\}$ (\textit{full-information outcome}).
    \item if $D \cup E_0 \neq \emptyset$\footnote{There are players who are not of the same type as the receiver.} and $A^* < E^*$, the equilibrium is characterized by \textit{no-information} outcome.
    \item if $D \cup E_0 \neq \emptyset$ and $A^* > E^*$, the equilibrium outcome $\tau^*$ is characterized by $supp(\tau^*) = \{\mu_{A^*},a\}$ where $a = a^*(p)$ if receiver is a conformist and $a = 1 - a^*(p)$ if he is a contrarian\footnote{Note that Assumption 1 is needed here if $A^* = R$; otherwise, there would be no equilibrium.}.
\end{enumerate}

\vspace{2mm}

\textit{In the equilibrium, if all players are of the same type as receiver, he gets full information. Otherwise\footnote{This is the case except for some rather uninteresting cases resulting in no information for the receiver, as explained in what follows.}, location of the pivotal player $A^*$ determines whether receiver gets any information, and if so, her bias $\mu_{A^*}$ determines the amount of information communicated to the receiver; the higher the bias of $A^*$, the more information communicated.}

\vspace{2mm}

\textbf{Proof}: The proof is along the same lines as that of Proposition 4 presented in Appendix A. \hfill$\blacksquare$

\vspace{2mm}

Another specific player will be of importance in the next results: Let $D^*$ represent the player with the lowest bias among those in $D$ with higher bias than $A^*$. If $\{i \in D: i\ \text{has higher bias than}\ A^*\} = \emptyset$, let $\mu_{D^*} = 1 - a^*(p)$ if receiver is a conformist and $\mu_{D^*} = a^*(p)$ if he is a contrarian.

\vspace{2mm}

No-information equilibrium outcome could emerge because of three reasons:
\begin{enumerate}
    \item Existence of players who would provide no information in a single-sender game: This is the case if there exists an extremist of the same bias as receiver or a player of opposite type to receiver but with the same bias or lower opposite bias $\mu_R < \mu_i < 1$, i.e., $C \cup E_1 \neq \emptyset$.
    \item Existence of players with highly opposed preferences: Assuming $C \cup E_1 = \emptyset$, this is the case if there exists $i \in D$ such that $i$ is less biased than $A^*$.
    \item Location of the pivotal player: Assuming $C \cup E_1 = \emptyset$ and there exists no $i \in D$ such that $i$ is more biased than $A^*$, this is the case if $A^* < E^*$, i.e., $E^*$ is closer to the receiver than the pivotal player $P$. This equilibrium is inefficient since all players prefer every outcome $\tau$ with $\min(supp(\tau)) \in [\mu_{D^*},\mu_{A^*}]$ or $\max(supp(\tau)) \in [\mu_{A^*},\mu_{D^*}]$ whichever is a well-defined interval. The equilibrium is inefficient if and only if this condition holds.
\end{enumerate}

\vspace{2mm}

\textbf{Corollary 5}: Let $\Omega = \mathcal{A} = \{0,1\}$ and suppose receiver is not an extremist and Assumption 1 holds. The equilibrium is inefficient if and only if the conditions of Proposition 5.b. hold.

\vspace{2mm}

\textbf{Corollary 6}: Let $\Omega = \mathcal{A} = \{0,1\}$ and suppose receiver is not an extremist and Assumption 1 holds. If receiver could add a player of his choice at the end of the hierarchy, he would choose one of the same type and bias as himself but with higher bias than $A^*$ and lower bias than $D^*$. Receiver's utility would then be increasing in the bias of the added player and the new equilibrium would be efficient. Moreover, he does not benefit from adding more players.

\subsection{General Binary-Action Games}

Let $\Omega = [0,1]$. Given the experiment $\pi_1$ chosen by player 1, Proposition 5 characterizes the equilibrium outcome of the binary-state, binary-action subgame starting from player 2. By choosing $\pi_1$, player 1 not only chooses the state space of the following subgame $supp(\tau_1) = \{q_0, q_1\}$, but also implicitly the four numbers representing the utilities of the subsequent players\footnote{$u_i^{lk} = \mathbb{E}_{q_l}[u_i(\omega, k)],\ \forall l,k \in \{0,1\}$}, and thus their types.

In the general binary-action game, for each player, there is a utility function corresponding to each action: $u_i(\omega, 0)$ and $u_i(\omega, 1)$. Let $\Delta u_i(\omega) = u_i(\omega,1) - u_i(\omega,0)$ and suppose $\Delta u_i(\omega) = \alpha_i \omega + \beta_i$ for some $\alpha_i,\beta_i \in \mathbb{R}$. For example, it could be the case that $u_i(\omega,0) = 0$ and $u_i(\omega,1) = \alpha_i \omega + \beta_i$. Let $\omega_i = - \frac{\beta_i}{\alpha_i}$ represent the state, or equivalently, the posterior mean, at which $i$ is indifferent between the two actions.
    
Generally, as mentioned before, player 1 chooses a Bayes-plausible distribution of posteriors $\tau_1 \in \Delta(\Delta(\Omega))$ with $supp(\tau_1) = \{q_0, q_1\}$. From this point on, the game is a binary-state, binary-action one. In the special case where $\Delta u_i(\omega) = \alpha_i \omega + \beta_i$, since action preferences only depend on the posterior mean, we can assume, with some abuse of notation, player 1 simply chooses a distribution of posterior means $\tau_1 \in \Delta(\Omega)$\footnote{Since $\tau_1$ is a finite-support distribution, $\tau_1$ represents the probability mass function.} where $supp(\tau_1) = \{m_0, m_1\}$\footnote{Recall that in the binary-state, binary-action games, each player was represented by the posterior belief $\mu_i$ (or equivalently, posterior mean) at which she was indifferent between the two actions. In this special case, in the binary-state, binary-action game starting from player 2, we have $\mu_i = \frac{\omega_i - m_0}{m_1 - m_0}$. Clearly, $\mu_i > \mu_j$ if and only if $\omega_i > \omega_j$.}, and $\tau_1$ is a mean preserving contraction of $F$ (implying $\mathbb{E}_{\tau_1}[\omega] = \mathbb{E}_F[\omega] = m$)\footnote{Clearly, $0 \leq m_0 \leq m \leq m_1 \leq 1$. In the special case where $\Delta u_i(\omega) = \alpha_i \omega + \beta_i$, with some abuse of notation, the outcome $\tau$ represents a distribution of posterior means.}. For example, if $F$ is the uniform distribution over $[0,1]$, a distribution of posterior means $\tau$ with $supp(\tau) = \{m_0, m_1\}$ gives a mean preserving contraction of  $F$ (with the Bayes-plausible probabilities, i.e., $\tau(m_0) m_0 + (1-\tau(m_0)) m_1 = m$) if and only if $m_1 - m_0 \leq 0.5$\footnote{Note that, generally, if the pair $(m_0, m_1)$ gives a mean preserving contraction of $F$, any pair $(m'_0, m'_1)$ also gives a mean preserving contraction of $F$ if and only if $m_0 \leq m'_0 \leq m \leq m'_1 \leq m_1$.}.
    
Generally, shape of the utility functions $u_i$ and choice of the first player $\tau_1$ determine the type of each subsequent player (conformist, contrarian, or extremist) in the following binary-state, binary-action subgame. In the special case where $\Delta u_i(\omega) = \alpha_i \omega + \beta_i$, all we need to know about player $i = 2,\ldots,n$ is her indifference posterior mean $\omega_i$\footnote{For simplicity, I assume $\omega_i \not\in \{0,m,1\}$ where $m = \mathbb{E}_F[\omega]$ and $\omega_i \neq \omega_j$ if $i \neq j$.} and the sign of $\alpha_i$ (or equivalently, her preferred action at $\omega = 1$)\footnote{This is similar to the claim we made in the binary-state, binary-action case since the subgame starting from player 2 is a binary-state, binary action one.}. If $\omega_i < 0$ or $\omega_i > 1$, player $i$ is called an absolute extremist; that is, regardless of $\tau_1$ chosen by player 1, she will be an extremist in the subgame starting from player 2. Otherwise, $\alpha_i > 0$ ($< 0$) implies that player $i$ is a conformist (contrarian). However, given $\tau_1$, and thus, $m_0$ and $m_1$, each conformist or contrarian may turn into an extremist in the subgame starting from player 2 as shown in Figure 11: this happens if $\omega_i < m_0$ or $\omega_i > m_1$; also, whenever $\tau_1$ turns an $a$-biased\footnote{Bias is defined similar to the binary-state, binary-action case: A player is biased toward action $a$ if she prefers that action given the common prior mean $m$.} conformist or contrarian into an extremist, she will become an $a$-extremist. The new extremists are related as follows: if player $i$ with $\omega_i < m$ ($\omega_i > m$) turns into an extremist, all players $j$ with $\omega_j < \omega_i$ ($\omega_j > \omega_i$) turn into extremists as well.

\begin{figure}[!h]
\centering
\subfloat[\small{Player $i$ remains a conformist with $\mu_i = \frac{\omega_i - m_0}{m_1 - m_0}$.}]{
    \begin{tikzpicture}
    \draw[blue, very thick] (-4,-5) -- (-2,-5);
    \draw[red, very thick] (-2,-5) -- (4,-5);
    
    \filldraw[black] (-4,-5) circle (2pt);
    \coordinate [label=below:$0$] (A) at (-4,-5.2);
    
    \filldraw[black] (4,-5) circle (2pt);
    \coordinate [label=below:$1$] (B) at (4,-5.2);
    
    \filldraw[black] (0,-5) circle (2pt);
    \coordinate [label=below:$m$] (C) at (0,-5.2);
    
    \filldraw[black] (-2,-5) circle (2pt);
    \coordinate [label=below:$\omega_i$] (D) at (-2,-5.2);
    
    \filldraw[green] (-3,-5) circle (2pt);
    \coordinate [label=below:$m_0$] (E) at (-3,-5.2);
    
    \filldraw[green] (3,-5) circle (2pt);
    \coordinate [label=below:$m_1$] (F) at (3,-5.2);
    
    \draw [decorate, decoration = {brace}, thin] (-4,-4.7) --  (-2.1,-4.7);
    \node[blue] at (-3,-4.2) {prefers $a = 0$};
    
    \draw [decorate, decoration = {brace}, thin] (-1.9,-4.7) --  (4,-4.7);
    \node[red] at (1,-4.2) {prefers $a = 1$};
    
    \end{tikzpicture}
}

\subfloat[\small{Player $i$ turns into a 1-extremist.}]{
    \begin{tikzpicture}
    \draw[blue, very thick] (-4,-5) -- (-2,-5);
    \draw[red, very thick] (-2,-5) -- (4,-5);
    
    \filldraw[black] (-4,-5) circle (2pt);
    \coordinate [label=below:$0$] (A) at (-4,-5.2);
    
    \filldraw[black] (4,-5) circle (2pt);
    \coordinate [label=below:$1$] (B) at (4,-5.2);
    
    \filldraw[black] (0,-5) circle (2pt);
    \coordinate [label=below:$m$] (C) at (0,-5.2);
    
    \filldraw[black] (-2,-5) circle (2pt);
    \coordinate [label=below:$\omega_i$] (D) at (-2,-5.2);
    
    \filldraw[green] (-1,-5) circle (2pt);
    \coordinate [label=below:$m_0$] (E) at (-1,-5.2);
    
    \filldraw[green] (3,-5) circle (2pt);
    \coordinate [label=below:$m_1$] (F) at (3,-5.2);
    
    \draw [decorate, decoration = {brace}, thin] (-4,-4.7) --  (-2.1,-4.7);
    \node[blue] at (-3,-4.2) {prefers $a = 0$};
    
    \draw [decorate, decoration = {brace}, thin] (-1.9,-4.7) --  (4,-4.7);
    \node[red] at (1,-4.2) {prefers $a = 1$};
    
    \end{tikzpicture}
}
\caption{The distribution of posterior means $\tau_1$ chosen by player 1 with $supp(\tau_1) = \{m_0,m_1\}$ may turn a conformist biased toward $a = 1$ into a 1-extremist in the subgame starting from player 2.}
\end{figure}
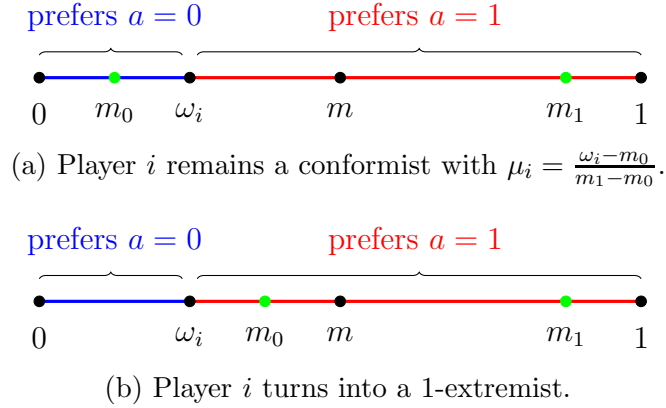



Consider the binary-state, binary-action game starting from player 2 where $\Omega = \{0,1\}$ and $\mu_i = \omega_i$, for all $i = 2,\ldots,n, R$\footnote{Note that the only extremists of this game are the absolute extremists of the original game.}. We call this game the \textit{reduced} binary-state game corresponding to the general binary-action game. 

\vspace{2mm}

\textbf{Lemma 1}: Let $\Delta u_i(\omega) = \alpha_i \omega + \beta_i$ for $i = 2,\ldots,n,R$. If the equilibrium outcome of the reduced binary-state game corresponding to the binary-action game is the no-information outcome, the same holds for the original binary-action game\footnote{Lemma 1 can be written more generally as follows: If the equilibrium outcome of the reduced binary-state game starting from player $i \geq 2$ is the no-information outcome, the same holds for the binary-action game starting from player $j < i$.}.  

\vspace{2mm}

\textbf{Proof}: See Appendix B. \hfill$\blacksquare$
    
\vspace{2mm}    
    
In general, the optimal choice of player 1 is not clear just based on $\Delta u_1(\omega)$. To make analysis tractable, let's assume $u_1(\omega,0) = 0$\footnote{This implies that $u_1(\omega,1) = \alpha_1 \omega + \beta_1$.}. Consider the binary-state, binary-action game where $\Omega = \{0,1\}$ and $\mu_i = \omega_i$, for all $i = 1,\ldots,n, R$. We call this game the binary-state game corresponding to the general binary-action game.

\vspace{2mm}

\textbf{Lemma 2}: Let $\Delta u_i(\omega) = \alpha_i \omega + \beta_i$ for $i = 1,\ldots,n,R$ and $u_1(\omega,0) = 0$. If the equilibrium outcome of the binary-state game corresponding to the binary-action game is the no-information outcome, the same holds for the original binary-action game\footnote{Lemma 2 can be written more generally as follows: If the equilibrium outcome of the binary-state game starting from player $i$ with $u_i(\omega,0) = 0$ is the no-information outcome, the same holds for the binary-action game starting from player $j \leq i$.}.

\vspace{2mm}

\textbf{Proof}: See Appendix B. \hfill$\blacksquare$

\vspace{2mm}

Again, if receiver is a an absolute extremist, he will always take action $a = 0$ or $a = 1$ regardless of the experiments chosen by the players, and thus all players are indifferent among all experiments. Given Assumption 2, receiver gets full information in equilibrium\footnote{This is clearly true for any binary-action game regardless of the assumptions we have made in this section.}.
    
Suppose receiver is a conformist who is biased toward $a = 1$, i.e., $\alpha_R > 0$ and $0< \omega_R < m$. Define the sets of players (not including player 1) $A,B,C,D,E_0,E_1$ as in Proposition 4 by replacing $\mu_i$ and $\mu_R$ with $\omega_i$ and $\omega_R$, respectively. Similarly, define $A^*$ and $E^*$.

The next proposition implies that the relative location and relative bias of three specific players in addition to player 1 and $E^*$ are relevant in determining the equilibrium outcome of the binary-action game in the special case where $u_1(\omega,0) = 0$ and the prior belief $F$ is uniform:

\begin{itemize}
    \item Let $P$ represent the conformist with the highest bias toward $a = 1$\footnote{This implies $P = 1$ or $P = A^*$.}: $\omega_P = \min_{i: \alpha_i > 0, 0 < \omega_i < 1} \omega_i$.
    \item Let $B^*$ represent the player with the highest bias among those in $B$: $\omega_{B^*} = \max_{i \in B} \omega_i$. If $B = \emptyset$, let $\omega_{B^*} = m$.
    \item Let $B^*_P$ represent the player with the highest bias among those in $B$ and closer to the receiver than $P$: $\omega_{B^*_P} = \max_{i \in B: i > P} \omega_i$. If $\{i \in B: i > P\} = \emptyset$, let $\omega_{B^*_P} = m$.
\end{itemize}

Figure 12 illustrates an example.

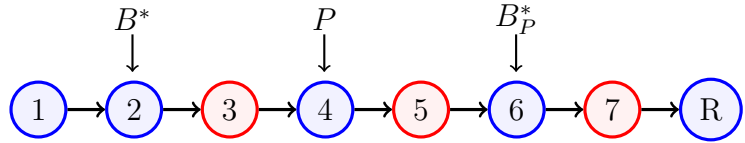
\begin{figure}[!h]
\centering
\begin{tikzpicture}[
conformist/.style={circle, draw=blue, fill=blue!5, very thick, minimum size=7mm},
contrarian/.style={circle, draw=red , fill=red!5 , very thick, minimum size=7mm}, node distance=0.5cm]

\node[conformist]   (p1)                  {1};
\node[conformist]   (p2)    [right=of p1] {2};
\node[contrarian]   (p3)    [right=of p2] {3};
\node[conformist]   (p4)    [right=of p3] {4};
\node[contrarian]   (p5)    [right=of p4] {5};
\node[conformist]   (p6)    [right=of p5] {6};
\node[contrarian]   (p7)    [right=of p6] {7};
\node[conformist]   (p8)    [right=of p7] {R};

\draw[->, very thick] (p1.east) -- (p2.west);
\draw[->, very thick] (p2.east) -- (p3.west);
\draw[->, very thick] (p3.east) -- (p4.west);
\draw[->, very thick] (p4.east) -- (p5.west);
\draw[->, very thick] (p5.east) -- (p6.west);
\draw[->, very thick] (p6.east) -- (p7.west);
\draw[->, very thick] (p7.east) -- (p8.west);

\draw[->, thick] (1.25,1) -- (1.25,0.5);
\node[black] at (1.25,1.2) {$B^*$};

\draw[->, thick] (3.8,1) -- (3.8,0.5);
\node[black] at (3.8,1.2) {$P$};

\draw[->, thick] (6.35,1) -- (6.35,0.5);
\node[black] at (6.35,1.2) {$B^*_P$};
\end{tikzpicture}
\caption{The blue circles represent the conformists while the red ones represent extremists and contrarians of a hierarchy with $n = 7$.}
\end{figure}

\vspace{2mm}
    
\textbf{Proposition 6}: Let $\Omega = [0,1]$, $\mathcal{A} = \{0,1\}$, $\Delta u_i(\omega) = \alpha_i \omega + \beta_i$ for $i = 1,\ldots,n,R$, $u_1(\omega,0) = 0$ and the prior belief $F$ be the uniform distribution over $[0,1]$. Suppose receiver is a conformist who is biased toward $a = 1$\footnote{$\alpha_R > 0$ and $0< \omega_R < m$.}, Assumptions 1 and 2 hold and the no-information outcome is not the equilibrium outcome of the corresponding binary-state game. The equilibrium is characterized by \textit{no-information outcome}\footnote{No information outcome is characterized by $supp(\tau^*) \subset (\omega_R,1]$ if $a^*(\omega_R) = 0$ or $supp(\tau^*) \subset [\omega_R,1]$ if $a^*(\omega_R) = 1$, which is determined by Assumption 1. However, under the assumption that $\mathcal{S}_n = \mathcal{A}$, no-information outcome is equivalent to $supp(\tau^*) = {m}$.} unless $\omega_{B^*_P} - \omega_{P} \leq 0.5$\footnote{That is, there would be no new 0-extremists closer to the receiver than $P$ if information is provided: $\omega_i \leq \omega_{P}+0.5$ for all players $i > P$.}; in this case,
    \begin{enumerate}[label=\alph*.]
         \item if there are players who are not conformists\footnote{That is, there are players who are absolute 0-extremists or contrarians with $0 < \omega_i < \omega_P$.}, $supp(\tau^*) = \{\omega_P, \omega_P + 0.5\}$.
         \item if all players are conformists and $\omega_{B^*} - \omega_{P} > 0.5$, $supp(\tau^*) = \{\omega_P,\omega_P+0.5\}$.
        \item if all players are conformists and $\omega_{B^*} - \omega_{P} \leq 0.5$\footnote{That is, there would be no new 0-extremists if information is provided: $\omega_i \leq \omega_{P}+0.5$ for all players.}, $supp(\tau^*) = \{m_0,m_0+0.5\}$ where $m_0 = \min\{\omega_P, \max(\omega_{B^*} - 0.5, 0.5\omega_1)\}$\footnote{Equivalently, $m_0 = \max\{\omega_{B^*} - 0.5, \min(\omega_P, 0.5\omega_1)\}$.}.
    \end{enumerate}
    
\vspace{2mm}    
    
\textit{Assuming the equilibrium outcome of the corresponding binary-state game is not the no-information outcome, receiver gets some information if and only if $\omega_{B^*_P} - \omega_P \leq 0.5$, i.e., there is not a conformist whose preferences are somewhat opposed to those of the pivotal player $P$ and is closer to the receiver. In this case, the biases of player 1 and the pivotal players $P$ and $B^*$ determine the information communicated to the receiver as shown in Figure 13. Every equilibrium is outcome-equivalent to one which simply distinguishes between the two intervals $[0,x]$ and $[x,1]$ for some $x \in [0,\omega_P]$.} 

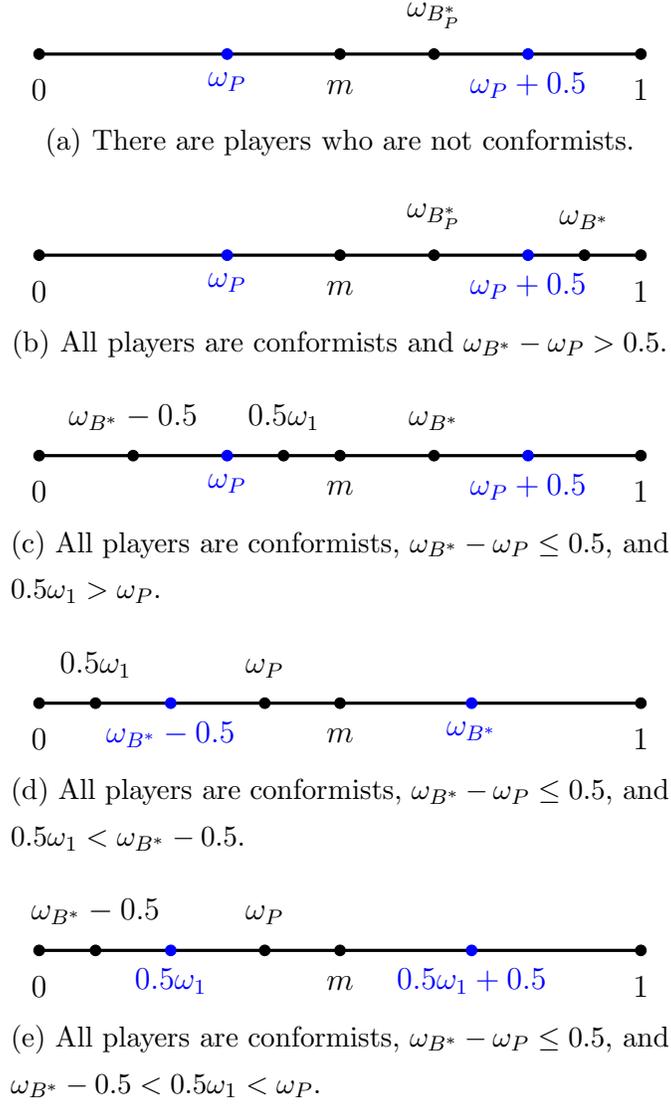
\begin{figure}[!h]
\centering
\subfloat[\small{There are players who are not conformists.}]{
\begin{tikzpicture}
    \draw[black, very thick] (-4,-5) -- (4,-5);
    
    \filldraw[black] (-4,-5) circle (2pt);
    \coordinate [label=below:$0$] (A) at (-4,-5.2);
    
    \filldraw[black] (4,-5) circle (2pt);
    \coordinate [label=below:$1$] (B) at (4,-5.2);
    
    \filldraw[black] (0,-5) circle (2pt);
    \coordinate [label=below:$m$] (C) at (0,-5.2);
    
    \filldraw[black] (1.25,-5) circle (2pt);
    \coordinate [label=above:$\omega_{B^*_P}$] (D) at (1.25,-4.8);
    
    \filldraw[blue] (-1.5,-5) circle (2pt);
    \coordinate [label={[blue]below:$\omega_P$}] (D) at (-1.5,-5.1);
    
    \filldraw[blue] (2.5,-5) circle (2pt);
    \coordinate [label={[blue]below:$\omega_P + 0.5$}] (D) at (2.5,-5.1);
\end{tikzpicture}
}

\subfloat[\small{All players are conformists and $\omega_{B^*} - \omega_P > 0.5$.}]{
\begin{tikzpicture}
    \draw[black, very thick] (-4,-5) -- (4,-5);
    
    \filldraw[black] (-4,-5) circle (2pt);
    \coordinate [label=below:$0$] (A) at (-4,-5.2);
    
    \filldraw[black] (4,-5) circle (2pt);
    \coordinate [label={[black]below:$1$}] (B) at (4,-5.2);
    
    \filldraw[black] (0,-5) circle (2pt);
    \coordinate [label=below:$m$] (C) at (0,-5.2);
    
    \filldraw[black] (3.25,-5) circle (2pt);
    \coordinate [label=above:$\omega_{B^*}$] (C) at (3.25,-4.8);
    
    \filldraw[black] (1.25,-5) circle (2pt);
    \coordinate [label=above:$\omega_{B^*_P}$] (C) at (1.25,-4.8);
    
    \filldraw[blue] (-1.5,-5) circle (2pt);
    \coordinate [label={[blue]below:$\omega_P$}] (D) at (-1.5,-5.1);
    
    \filldraw[blue] (2.5,-5) circle (2pt);
    \coordinate [label={[blue]below:$\omega_P + 0.5$}] (D) at (2.5,-5.1);
\end{tikzpicture}
}

\subfloat[\small{All players are conformists, $\omega_{B^*} - \omega_P \leq 0.5$, and $0.5 \omega_1 > \omega_P$.}]{
\begin{tikzpicture}
    \draw[black, very thick] (-4,-5) -- (4,-5);
    
    \filldraw[black] (-4,-5) circle (2pt);
    \coordinate [label=below:$0$] (A) at (-4,-5.2);
    
    \filldraw[black] (4,-5) circle (2pt);
    \coordinate [label={[black]below:$1$}] (B) at (4,-5.2);
    
    \filldraw[black] (0,-5) circle (2pt);
    \coordinate [label=below:$m$] (C) at (0,-5.2);
    
    \filldraw[black] (1.25,-5) circle (2pt);
    \coordinate [label=above:$\omega_{B^*}$] (C) at (1.25,-4.8);
    
    \filldraw[black] (-2.75,-5) circle (2pt);
    \coordinate [label=above:$\omega_{B^*} - 0.5$] (C) at (-2.75,-4.8);
    
    \filldraw[black] (-0.75,-5) circle (2pt);
    \coordinate [label=above:$0.5 \omega_1$] (C) at (-0.75,-4.8);
    
    \filldraw[blue] (-1.5,-5) circle (2pt);
    \coordinate [label={[blue]below:$\omega_P$}] (D) at (-1.5,-5.1);
    
    \filldraw[blue] (2.5,-5) circle (2pt);
    \coordinate [label={[blue]below:$\omega_P + 0.5$}] (D) at (2.5,-5.1);
\end{tikzpicture}
}

\subfloat[\small{All players are conformists, $\omega_{B^*} - \omega_P \leq 0.5$, and $0.5 \omega_1 < \omega_{B^*} - 0.5$.}]{
\begin{tikzpicture}
\draw[black, very thick] (-4,-5) -- (4,-5);
    
    \filldraw[black] (-4,-5) circle (2pt);
    \coordinate [label=below:$0$] (A) at (-4,-5.2);
    
    \filldraw[black] (4,-5) circle (2pt);
    \coordinate [label={[black]below:$1$}] (B) at (4,-5.2);
    
    \filldraw[black] (0,-5) circle (2pt);
    \coordinate [label=below:$m$] (C) at (0,-5.2);
    
    \filldraw[blue] (1.75,-5) circle (2pt);
    \coordinate [label={[blue]below:$\omega_{B^*}$}] (C) at (1.75,-5.1);
    
    \filldraw[blue] (-2.25,-5) circle (2pt);
    \coordinate [label={[blue]below:$\omega_{B^*} - 0.5$}] (C) at (-2.25,-5.1);
    
    \filldraw[black] (-3.25,-5) circle (2pt);
    \coordinate [label=above:$0.5 \omega_1$] (C) at (-3.25,-4.8);
    
    \filldraw[black] (-1,-5) circle (2pt);
    \coordinate [label=above:$\omega_P$] (D) at (-1,-4.8);
    
\end{tikzpicture}
}

\subfloat[\small{All players are conformists, $\omega_{B^*} - \omega_P \leq 0.5$, and $\omega_{B^*} - 0.5 < 0.5 \omega_1 < \omega_P$.}]{
\begin{tikzpicture}
\draw[black, very thick] (-4,-5) -- (4,-5);
    
    \filldraw[black] (-4,-5) circle (2pt);
    \coordinate [label=below:$0$] (A) at (-4,-5.2);
    
    \filldraw[black] (4,-5) circle (2pt);
    \coordinate [label={[black]below:$1$}] (B) at (4,-5.2);
    
    \filldraw[black] (0,-5) circle (2pt);
    \coordinate [label=below:$m$] (C) at (0,-5.2);
    
    \filldraw[blue] (1.75,-5) circle (2pt);
    \coordinate [label={[blue]below:$0.5 \omega_1 + 0.5$}] (C) at (1.75,-5.1);
    
    \filldraw[blue] (-2.25,-5) circle (2pt);
    \coordinate [label={[blue]below:$0.5 \omega_1$}] (C) at (-2.25,-5.1);
    
    \filldraw[black] (-3.25,-5) circle (2pt);
    \coordinate [label=above:$\omega_{B^*} - 0.5$] (C) at (-3.25,-4.8);
    
    \filldraw[black] (-1,-5) circle (2pt);
    \coordinate [label=above:$\omega_P$] (D) at (-1,-4.8);
    
\end{tikzpicture}
}

\caption{The equilibrium outcome if $\omega_{B^*_P} - \omega_P \leq 0.5$: The blue dots represent the support of the equilibrium distribution of posteriors.}
\end{figure}

\vspace{2mm}
    
\textbf{Proof}: We first state two lemmas which are proved in Appendix B:

\vspace{2mm}

\textit{Lemma 3}: Let $\Omega = [0,1]$, $\mathcal{A} = \{0,1\}$ and $\Delta u_i(\omega) = \alpha_i \omega + \beta_i$ for $i = 2,\ldots,n,R$. Suppose receiver is a conformist who is biased toward $a= 1$, Assumptions 1 and 2 hold and the no-information outcome is not the equilibrium outcome of the reduced binary-state game\footnote{This implies that (not including player 1) there exists no absolute 1-extremists or contrarians with $\omega_{A^*} < \omega_i < 1$, and $A^*$ is closer to the receiver than all absolute 0-extremists and contrarians with $0 < \omega_i < \omega_{A^*}$.}. Given player 1's choice of distribution of posterior means $\tau_1$ with $supp(\tau_1) = \{m_0, m_1\}$, the equilibrium of the following subgame starting from player 2 is characterized by \textit{no-information} outcome unless:
\begin{enumerate}[label=\alph*.]
    \item (i) all remaining players are conformists\footnote{$C \cup D \cup E_0 \cup E_1 = \emptyset$.}, (ii) $m_0 \leq \omega_{A^*}$ and $m_1 \geq \omega_{B^*}$: In this case, the equilibrium outcome $\tau^*$ of the following subgame starting from player 2 is characterized by $supp(\tau^*) = \{m_0, m_1\}$ (\textit{full-information outcome})\footnote{Without Assumption 2, if $A^* \neq R$ and $m_0 = \omega_{A^*}$, another possible equilibrium is no-information outcome. Similarly, if $m_1 = \omega_{B^*}$, in other possible equilibria, player 1's choice of distribution of posterior means ($m_0, m_1$) would reach the receiver as $(m', \omega_{B^*})$ where $m_0 \leq m' \leq \omega_{A^*}$.}. 
    \item (i) all remaining players are conformists but $m_1 < \omega_{B^*}$, or there are players (not including player 1) who are not conformists, (ii) $m_0 \leq \omega_{A^*}$, (iii) $A^*$ is closer to the receiver than all the new 0-extremists, i.e., conformists with $\omega_i > m_1$: In this case, the equilibrium outcome $\tau^*$ of the following subgame starting from player 2 is characterized by $supp(\tau^*) = \{\mu_{A^*} m_1 + (1-\mu_{A^*}) m_0 = \omega_{A^*}, m_1\}$\footnote{Note that Assumption 1 is needed here if $A^* = R$; otherwise, there would be no equilibrium.}.
\end{enumerate}

\vspace{2mm}

\textit{Lemma 4}: Let $\Omega = [0,1]$, $\mathcal{A} = \{0,1\}$, $\Delta u_i(\omega) = \alpha_i \omega + \beta_i$ for $i = 1,\ldots,n,R$ and $u_1(\omega,0) = 0$. Suppose receiver is a conformist who is biased toward $a= 1$ and Assumption 2 holds. The preferences of different types of player 1 over outcomes $\tau$ with $supp(\tau) = \{m_0,m_1\}$ are as follows:
    \begin{itemize}
        \item If player 1 is an absolute 1-extremist or a contrarian ($\alpha_1 < 0$) with $\omega_R < \omega_1 < 1$, she would prefer the no-information outcome to all other outcomes\footnote{Thus, the equilibrium outcome $\tau^*$ is characterized by $supp(\tau^*) \subset (\mu_R,1]$ if $a^*(\mu_R) = 0$ or $supp(\tau^*) \subset [\mu_R,1]$ if $a^*(\mu_R) = 1$, which is determined by Assumption 1 (\textit{no-information outcome}). However, under the assumption that $\mathcal{S}_n = \mathcal{A}$, no-information outcome is equivalent to $supp(\tau^*) = {m}$. Note that $u_1(\omega,0) = 0$ is not necessary for this part.}. 
        \item If player 1 is an absolute 0-extremist, she would prefer outcomes with higher $m_1$ and higher $m_0$ as long as it is not the no-information outcome.
        \item If player 1 is a contrarian ($\alpha_1 < 0$) with $0 < \omega_1 < \omega_R$, she would prefer outcomes with higher $m_0$ and $m_1$, as long as $m_0 \geq \omega_1$ and it is not the no-information outcome\footnote{In this case, if $m_0 = \omega_1$, she would be indifferent about $m_1$ and Assumption 2 implies this part.}. She would also prefer the no-information outcome to all outcomes with $m_0 < \omega_1$.
        \item If player 1 is a conformist with $0 < \omega_1 < \omega_R$, she would prefer outcomes with lower $m_0$ and higher $m_1$ as long as $m_0 \leq \omega_1$\footnote{In this case, if $m_0 = \omega_1$, she would be indifferent about $m_1$ and Assumption 2 implies this part.}. She would also prefer the no-information outcome only to outcomes with $m_0 > \omega_1$.
        \item If player 1 is a conformist ($\alpha_1 > 0$) with $\omega_R < \omega_1 < m$, she would prefer lower $m_0$ and higher $m_1$ as long as it is not the no-information outcome.
        \item If player 1 is a conformist ($\alpha_1 > 0$) with $m < \omega_1 < 1$, she would prefer higher $m_1$ and (i) lower $m_0$ if $m_1 \geq \omega_1$, (ii) higher $m_0$ if $m_1 < \omega_1$, as long as it is not the no-information outcome\footnote{In this case, if $m_1 = \omega_1$, she would be indifferent among all $m_0$ as long as it is not the no-information outcome and Assumption 2 implies this part.}.
    \end{itemize}
    
\vspace{2mm}    

Now, we can prove Proposition 6. Note that there are no absolute 1-extremists or contrarians with $\omega_P < \omega_i < 1$ and $P > E^*$. As mentioned before, if $F$ is the uniform distribution over $[0,1]$, a distribution of posterior means $\tau$ with $supp(\tau) = \{m_0, m_1\}$ gives a mean preserving contraction of  $F$ (with the Bayes-plausible probabilities, i.e., $\tau(m_0) m_0 + (1-\tau(m_0)) m_1 = m$) if and only if $m_1 - m_0 \leq 0.5$. 

Suppose $\omega_{B^*_P} - \omega_{P} > 0.5$.
\begin{itemize}
    \item If $P = A^* = R$: Every possible choice of $m_0$ and $m_1$ by player 1 either is the no-information outcome or turns $B^*_P$ into a 0-extremist which leads to no-information outcome by Lemma 3.  
    \item If $P = 1$: Every possible choice of $m_0$ and $m_1$ by player 1 with $m_0 \leq \omega_1$ turns $B^*_P$ into a 0-extremist which leads to no-information outcome by Lemma 3. Moreover, by Lemma 4, she would prefer the no-information outcome to every possible choice of $m_0$ and $m_1$ with $m_0 > \omega_1$ which is not the no-information outcome. Therefore, the only possible equilibrium outcome is the no-information outcome.
    \item If $P = A^* \neq 1$: Every possible choice of $m_0$ and $m_1$ by player 1 either turns $P$ into a 1-extremist or turns $B^*_P$ into a 0-extremist, both of which lead to no-information outcome by Lemma 3.
\end{itemize}

Suppose $\omega_{B^*_P} - \omega_{P} \leq 0.5$. Before proceeding, note that given any outcome (other than the no-information outcome) with $m_1 = m_0 + 0.5$, the expected utility of player 1 is given by
\begin{align*}
    \mathbb{E}_{\tau}[u_1(\omega,a^*(\omega)] &= \frac{m - m_0}{m_1 - m_0} (\alpha_1 m_1 + \beta_1)\\
    &= 2(0.5 - m_0) (\alpha_1 m_0 + 0.5 \alpha_1 + \beta_1)
\end{align*}
where we have used the fact that $\mathbb{E}_F[\omega] = m$. Taking the derivative 
$$\frac{\partial \mathbb{E}_{\tau}[u_1(\omega,a^*(\omega)]}{\partial m_0} = -2 \alpha_1 m_0 - \beta_1$$
we see that it is equal to zero at $m_0 = \frac{-\beta_1}{2 \alpha_1} = 0.5 \omega_1$ where the expected utility of player 1 is maximized if she is a conformist ($\alpha_1 > 0$) and minimized if she is a contrarian ($\alpha_1 < 0$). Moreover, the expected utility of player 1 is increasing in $m_0$ if she is a 0-extremist. 
\begin{itemize}
    \item Suppose there are players who are not conformists, that is, there are players who are absolute 0-extremists or contrarians with $0 < \omega_i < \omega_P$. Note that $P \neq 1$ since otherwise, $P < E^*$ and no-information outcome is the equilibrium outcome of the corresponding binary-state game. 
    
    Suppose the only non-conformist is player 1. By Lemma 3, every possible choice of $m_0$ and $m_1$ by player 1 with $m_0 \leq \omega_P$ and $m_1 \geq \omega_{B^*}$ leads to the outcome $\tau$ with $supp(\tau) = \{m_0, m_1\}$; since by Lemma 4, player 1 would prefer outcomes with higher $m_0$, she chooses $m_0 = \omega_P$. Similarly, every possible choice of $m_0$ and $m_1$ by player 1 with $m_0 \leq \omega_P$ and $\omega_{B^*_P} \leq m_1 < \omega_{B^*}$ leads to the outcome $\tau$ with $supp(\tau) = \{\omega_P, m_1\}$; all other possible choices of $m_0$ and $m_1$ by player 1 lead to no-information outcome. Since by Lemma 4, player 1 would prefer outcomes with higher $m_1$, she chooses $m_1 = m_0 + 0.5$. Therefore, the equilibrium outcome would be $m_0 = \omega_P$ and $m_1 = \omega_P + 0.5$ since $\omega_P + 0.5 \geq \omega_{B^*_P}$.
    
    Suppose there are non-conformists among players other than player 1. By Lemma 3, every possible choice of $m_0$ and $m_1$ by player 1 with $m_0 \leq \omega_P$ and $m_1 \geq \omega_{B^*_P}$ leads to the outcome $\tau$ with $supp(\tau) = \{\omega_P, m_1\}$; all other choices of $m_0$ and $m_1$ by player 1 lead to no-information outcome. Since by Lemma 4, player 1 would prefer outcomes with higher $m_1$, she chooses $m_1 = m_0 + 0.5$. Therefore, the equilibrium outcome would be $m_0 = \omega_P$ and $m_1 = \omega_P + 0.5$ since $\omega_P + 0.5 \geq \omega_{B^*_P}$.
    
    \item Suppose all players are conformists and $\omega_{B^*} - \omega_P > 0.5$. Every possible choice of $m_0$ and $m_1$ by player 1 with $m_0 \leq \omega_P$ turns $B^*$ into a 0-extremist, and by Lemma 3, leads to the outcome $\tau$ with $supp(\tau) = \{\omega_P, m_1\}$ as long as $m_1 \geq \omega_{B^*_P}$; all other possible choices of $m_0$ and $m_1$ by player 1 lead to no-information outcome. Since by Lemma 4, player 1 would prefer outcomes with higher $m_1$, she chooses $m_1 = m_0 + 0.5$. Therefore, the equilibrium outcome would be $m_0 = \omega_P$ and $m_1 = \omega_P + 0.5$ since $\omega_P + 0.5 \geq \omega_{B^*_P}$.
    
    \item Suppose all players are conformists and $\omega_{B^*} - \omega_P \leq 0.5$. By Lemma 3, every possible choice of $m_0$ and $m_1$ by player 1 with $m_0 \leq \omega_P$ and $m_1 \geq \omega_{B^*}$ leads to the outcome $\tau$ with $supp(\tau) = \{m_0, m_1\}$. Similarly, every possible choice of $m_0$ and $m_1$ by player 1 with $m_0 \leq \omega_P$ and $\omega_{B^*_P} \leq m_1 < \omega_{B^*}$ leads to the outcome $\tau$ with $supp(\tau) = \{\omega_P, m_1\}$; all other possible choices of $m_0$ and $m_1$ by player 1 lead to no-information outcome. Since by Lemma 4, player 1 would prefer outcomes with higher $m_1$, she chooses $m_1 = m_0 + 0.5$. Her expected utility is maximized at $m_0 = 0.5 \omega_1$ if $0.5 \omega_1 \leq \omega_P$ and $0.5 \omega_1 + 0.5 \geq \omega_{B^*}$; if the first one is violated, she chooses $m_0 = \omega_P$, and if the second one is violated, she chooses $m_0 = \omega_{B^*} - 0.5$ since her expected utility is decreasing in $m_0$ in the interval $[0.5 \omega_1, \omega_P]$. Therefore, the equilibrium outcome would be $m_0 = \min\{\omega_P, \max(\omega_{B^*} - 0.5, 0.5\omega_1)\}$ and $m_1 = m_0 + 0.5$.
\end{itemize} \hfill$\blacksquare$

\vspace{2mm}

Another specific player will be of importance in the next results: Let $D^{**}$ represent the player with the lowest bias among those in $D$ with $\omega_i < \omega_P$: $\omega_{D^{**}} = \max_{i \in D: \omega_i < \omega_P} \omega_i$. If $\{i \in D: \omega_i < \omega_P\} = \emptyset$, let $\omega_{D^{**}} = 0$. If player 1 is a contrarian with $0 < \omega_1 < \omega_P$, let $\omega_{D^*} = \max(\omega_{D^{**}},\omega_1)$; otherwise, let $\omega_{D^*} = \omega_{D^{**}}$.

\vspace{2mm}

No-information equilibrium outcome could emerge because of four reasons:

\begin{enumerate}
    \item \textit{Condition 1}: No-information outcome is the equilibrium outcome of the corresponding binary-state game.
    \begin{enumerate}
        \item Existence of players who would provide no information in a single-sender game: This is the case if there exists an absolute 1-extremist or a contrarian ($\alpha_1 < 0$) with $\omega_R < \omega_i < 1$.
        \item Existence of players with highly opposed preferences: If condition (a) does not hold, this is the case if there exists a contrarian ($\alpha_1 < 0$) with $\omega_P < \omega_i < \omega_R$.
        \item Location of the pivotal player $P$: If conditions (a) and (b) do not hold, this is the case if $P < E^*$, i.e., $E^*$ is closer to the receiver than the pivotal player $P$. This equilibrium is inefficient since all players prefer every outcome $\tau$ with $\min(supp(\tau)) \in [\omega_{D^{**}},\omega_P]$.
    \end{enumerate}
    \item \textit{Condition 2}: Location and bias of the pivotal player $P$: If condition (1) does not hold, this is the case if $\omega_{B^*_P} - \omega_P > 0.5$, i.e., there exists a conformist whose preferences are somewhat opposed to those of the pivotal player $P$ and is closer to the receiver as shown in Figure 14. This equilibrium is inefficient since all players prefer every outcome $\tau$ with $\min(supp(\tau)) \in [\omega_{D^{**}},\omega_P]$.
    
\begin{figure}[!h]
\centering
\subfloat[\small{If $\omega_{B^*_P} - \omega_P > 0.5$, either $m_0 > \omega_P$ or $m_1 < \omega_{B^*_P}$ for every choice of $m_0$ and $m_1$ such that $m_1 - m_0 \leq 0.5$.}]{
    \begin{tikzpicture}
    \draw[black, very thick] (-4,-5) -- (4,-5);
    
    \filldraw[black] (-4,-5) circle (2pt);
    \coordinate [label=below:$0$] (A) at (-4,-5.2);
    
    \filldraw[black] (4,-5) circle (2pt);
    \coordinate [label=below:$1$] (B) at (4,-5.2);
    
    \filldraw[black] (0,-5) circle (2pt);
    \coordinate [label=below:$m$] (C) at (0,-5.2);
    
    \filldraw[blue] (-2.25,-5) circle (2pt);
    \coordinate [label=below:$\omega_P$] (D) at (-2.25,-5.2);
    
    \filldraw[blue] (2.25,-5) circle (2pt);
    \coordinate [label=below:$\omega_{B^*_P}$] (E) at (2.25,-5.2);
    \end{tikzpicture}
}

\subfloat[\small{If $m_0 > \omega_P$, $P$ turns into a 1-extremist.}]{
    \begin{tikzpicture}
    \draw[black, very thick] (-4,-5) -- (4,-5);
    
    \filldraw[black] (-4,-5) circle (2pt);
    \coordinate [label=below:$0$] (A) at (-4,-5.2);
    
    \filldraw[black] (4,-5) circle (2pt);
    \coordinate [label=below:$1$] (B) at (4,-5.2);
    
    \filldraw[black] (0,-5) circle (2pt);
    \coordinate [label=below:$m$] (C) at (0,-5.2);
    
    \filldraw[blue] (-2.25,-5) circle (2pt);
    \coordinate [label=below:$\omega_P$] (D) at (-2.25,-5.2);
    
    \filldraw[blue] (2.25,-5) circle (2pt);
    \coordinate [label=below:$\omega_{B^*_P}$] (E) at (2.25,-5.2);
    
    \filldraw[red] (-1,-5) circle (2pt);
    \coordinate [label=below:$m_0$] (D) at (-1,-5.2);
    
    \filldraw[red] (3,-5) circle (2pt);
    \coordinate [label=below:$m_1$] (E) at (3,-5.2);
    \end{tikzpicture}
}

\subfloat[\small{If $m_1 < \omega_{B^*_P}$, $B^*_P$ turns into a 0-extremist closer to the receiver than $P$.}]{
    \begin{tikzpicture}
    \draw[black, very thick] (-4,-5) -- (4,-5);
    
    \filldraw[black] (-4,-5) circle (2pt);
    \coordinate [label=below:$0$] (A) at (-4,-5.2);
    
    \filldraw[black] (4,-5) circle (2pt);
    \coordinate [label=below:$1$] (B) at (4,-5.2);
    
    \filldraw[black] (0,-5) circle (2pt);
    \coordinate [label=below:$m$] (C) at (0,-5.2);
    
    \filldraw[blue] (-2.25,-5) circle (2pt);
    \coordinate [label=below:$\omega_P$] (D) at (-2.25,-5.2);
    
    \filldraw[blue] (2.25,-5) circle (2pt);
    \coordinate [label=below:$\omega_{B^*_P}$] (E) at (2.25,-5.2);
    
    \filldraw[red] (-3,-5) circle (2pt);
    \coordinate [label=below:$m_0$] (D) at (-3,-5.2);
    
    \filldraw[red] (1,-5) circle (2pt);
    \coordinate [label=below:$m_1$] (E) at (1,-5.2);

    \end{tikzpicture}
}

\caption{If $\omega_{B^*_P} - \omega_P > 0.5$, no-information equilibrium outcome emerges.}
\end{figure}
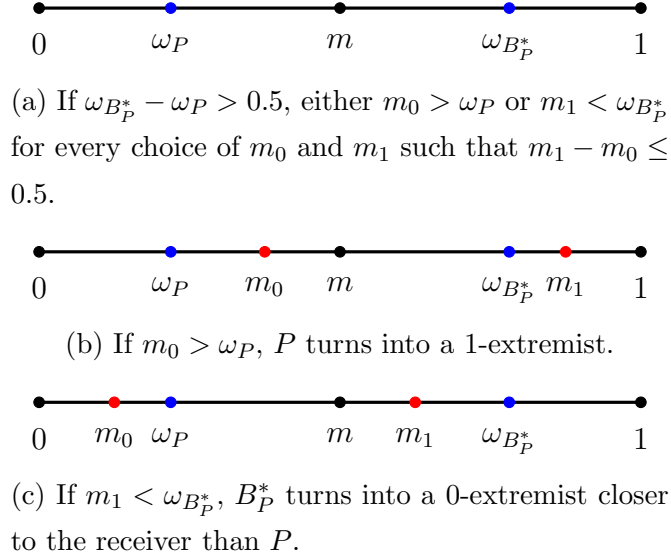
    
\end{enumerate}

Condition 1(c) and Condition 2 are similar: there exists a player with somewhat opposed preferences to those of the pivotal player $P$ who is closer to the receiver. The next corollary shows that this configuration is the only one which causes inefficiency in communication.

\vspace{2mm}

\textbf{Corollary 7}: Let $\Omega = [0,1]$, $\mathcal{A} = \{0,1\}$, $\Delta u_i(\omega) = \alpha_i \omega + \beta_i$ for $i = 1,\ldots,n,R$, $u_1(\omega,0) = 0$ and the prior belief $F$ be the uniform distribution over $[0,1]$. Suppose receiver is a conformist who is biased toward $a = 1$ and Assumptions 1 and 2 hold. The equilibrium is inefficient if and only if Condition 1(c) or Condition 2 holds.

\vspace{2mm}

Similar to the binary-state case, the reason is that when designing her experiment, $E^*$ or $B^*_P$, whose preferences are somewhat opposed to those of $P$, only considers the incentive compatibility constraints of the succeeding players, not those of the preceding players such as $P$. Knowing this, $P$ preemptively provides no information.

Similar to the binary-state case, as the next corollary shows, receiver can get more information and thus increase her payoff by assigning a suitable vice-president.

\vspace{2mm}

\textbf{Corollary 8}: Let $\Omega = [0,1]$, $\mathcal{A} = \{0,1\}$, $\Delta u_i(\omega) = \alpha_i \omega + \beta_i$ for $i = 1,\ldots,n,R$, $u_1(\omega,0) = 0$ and the prior belief $F$ be the uniform distribution over $[0,1]$. Suppose receiver is a conformist who is biased toward $a = 1$ and Assumptions 1 and 2 hold. If receiver could add a player of his choice at the end of the hierarchy, depending on the hierarchy configuration, he would either choose a conformist biased toward $a=1$ with $\omega_{n+1} = \min\{\omega_P,\max(0.5\omega_R,\omega_{D^{**}})\}$\footnote{Equivalently, $\omega_{n+1} = \max\{\omega_{D^{**}},\min(0.5\omega_R,\omega_P)\}$.}, or a conformist biased toward $a = 0$ with $\omega_{n+1} = \min(0.5\omega_R,\omega_P) + 0.5$. The new equilibrium would be efficient\footnote{To determine the new equilibrium outcome using Proposition 6, if there are two players with indifference belief $\omega_P$, we consider the closest one to the receiver as the pivotal player $P$.}.

\vspace{2mm}

If the players have sufficiently opposed preferences\footnote{That is, either all are not conformists or if they are, $\omega_{B^*} - \omega_P > 0.5$.}, similar to the binary-state case, every player is better off by respecting the preferences of the pivotal player $P$ as long as $P$ is closer to the receiver. By behaving as prescribed in the corollary receiver makes vice president the new pivotal player $P$, and all players are better off by respecting her preferences. Receiver chooses the bias of the vice president such that the outcome is as close as possible to the full-information outcome\footnote{Note that, from receiver's point of view, $supp(\tau) = \{0.5\omega_R,0.5\omega_R+0.5\}$ is equivalent to the full-information outcome.}. 

However, if all players have somewhat similar preferences\footnote{That is, all are conformists and $\omega_{B^*} - \omega_P \leq 0.5$.}, player 1 is better off by respecting the preferences of (i) not only the pivotal player $P$, (ii) but also the pivotal player $B^*$, as long as they are closer to the receiver. By behaving as prescribed in the corollary, receiver makes vice president the new pivotal player $P$ or $B^*$, and player 1 is better off by respecting her preferences. Receiver chooses the bias of the vice president such that the outcome is as close as possible to the full-information outcome. He follows (i) and chooses $\omega_{n+1} = \min(\omega_P, 0.5\omega_R)$ if $\min(supp(\tau^*)) > 0.5\omega_R$ or $\min(supp(\tau^*)) = \omega_P$; otherwise, he follows (ii) and chooses $\omega_{n+1} = \min(0.5\omega_R,\omega_P) + 0.5$.

\vspace{2mm}

\textbf{Corollary 9}: Let $\Omega = [0,1]$, $\mathcal{A} = \{0,1\}$, $\Delta u_i(\omega) = \alpha_i \omega + \beta_i$ for $i = 1,\ldots,n,R$, $u_1(\omega,0) = 0$ and the prior belief $F$ be the uniform distribution over $[0,1]$. Suppose receiver is a conformist who is biased toward $a = 1$ and Assumptions 1 and 2 hold. If receiver could add two players of his choice at the end of the hierarchy, he would choose a conformist biased toward $a=1$ with $\omega_{i} = \min\{\omega_P,\max(0.5\omega_R,\omega_{D^*})\}$ and a conformist biased toward $a = 0$ with $\omega_{n+1} = \min(0.5\omega_R,\omega_P) + 0.5$. The order does not matter and the new equilibrium would be efficient. Moreover, he does not benefit from adding more players.

\vspace{2mm}

Similar to the binary-state case, Proposition 6 can be generalized to other types of the receiver.

\section{Conclusion}
In this paper, we have investigated the outcome of intermediated communication in hierarchical organizations through the lens of Bayesian persuasion. We show that a version of revelation principle holds, that is, it is without loss to assume the only sender who may conceal information along the hierarchy is the initial one. We then use this simplification to show that hierarchical Bayesian persuasion is equivalent to single-sender Bayesian persuasion among the initial sender and the final receiver subject to recursively-defined incentive compatibility constraints dictated by the preferences of the intermediaries.

Applying these results to the case that the decision maker faces a binary decision underscores the importance of vice presidents in hierarchical organizations. We show that, in such cases, regardless of the number of intermediaries, a few of them are relevant in determining the outcome.

Given the above result, an interesting question that could be addressed in future research is whether we can still pinpoint a few relevant intermediaries if the decision is not binary. Another question to be addressed is the effect of private information on the outcome.

Moreover, this paper can be considered as a first step toward analyzing Bayesian persuasion in networks such as social media. The main difficulty of such an extension lies in analyzing the optimal experiment of a sender who is not aware of the prior of the receiving end.

\section{Appendix A: Binary-State, Binary-Action Games}

Bayes-plausibility implies that for any outcome of the game, or equivalently, any distribution of posteriors for the receiver $\tau$ with $supp(\tau) = \{q_0,q_1\}$, we have $\mathbb{E}_{\tau}[q] = p$. Without loss of generality, let $q_0 \leq q_1$ which implies $0 \leq q_0 \leq p \leq q_1 \leq 1$, $\tau(q_1) = \frac{p - q_0}{q_1 - q_0}$, and $\tau(q_0) = \frac{q_1 - p}{q_1 - q_0}$. Note that in the binary-state, binary-action case, being smaller in the convex order boils down to being a mean preserving contraction. The set of binary-support distributions of posteriors which are mean preserving contractions of $\tau$ includes all $\tau'$ with $supp(\tau') = \{q'_0,q'_1\}$ such that $q_0 \leq q'_0 \leq p \leq q'_1 \leq q_1$ and $\mathbb{E}_{\tau'}[q'] = p$. In other words, $\tau$ is more Blackwell-informative, the lower $q_0$ is and the higher $q_1$ is as shown in Figure 15.

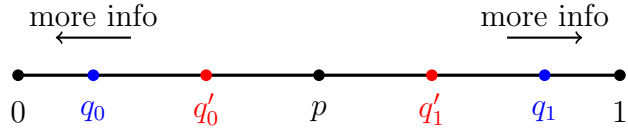
\begin{figure}[!h]
\centering
\begin{tikzpicture}
    \draw[black, very thick] (-4,-5) -- (4,-5);
    
    \filldraw[black] (-4,-5) circle (2pt);
    \coordinate [label=below:$0$] (A) at (-4,-5.2);
    
    \filldraw[black] (4,-5) circle (2pt);
    \coordinate [label=below:$1$] (B) at (4,-5.2);
    
    \filldraw[blue] (-3,-5) circle (2pt);
    \coordinate [label={[blue]below:$q_0$}] (C) at (-3,-5.2);
    
    \filldraw[red] (-1.5,-5) circle (2pt);
    \coordinate [label={[red]below:$q'_0$}] (D) at (-1.5,-5.1);
    
    \filldraw[black] (0,-5) circle (2pt);
    \coordinate [label=below:$p$] (E) at (0,-5.2);
    
    \filldraw[red] (1.5,-5) circle (2pt);
    \coordinate [label={[red]below:$q'_1$}] (F) at (1.5,-5.1);
    
    \filldraw[blue] (3,-5) circle (2pt);
    \coordinate [label={[blue]below:$q_1$}] (G) at (3,-5.2);
    
    \draw[->, thick] (-2.5,-4.5) -- (-3.5,-4.5);
    \node[black] at (-3,-4.2) {more info};
    
    \draw[->, thick] (2.5,-4.5) -- (3.5,-4.5);
    \node[black] at (3,-4.2) {more info};
    
    
\end{tikzpicture}
\caption{Comparison of outcomes}
\end{figure}

Suppose the receiver is a conformist who is biased toward $a = 1$. Also, for the ease of exposition, suppose receiver takes $a = 1$ at $\mu = \mu_R$. Given an outcome $\tau$ with $q_0 < \mu_R$, the expected utility of a player is given by
$$\mathbb{E}_{\tau}[v(q)] = \frac{p - q_0}{q_1 - q_0} [q_1 u^{11} + (1 - q_1) u^{01}] + \frac{q_1 - p}{q_1 - q_0} [q_0 u^{10} + (1 - q_0) u^{00}]$$
Moreover, given any outcome $\tau$ with $q_0 \geq \mu_R$, the expected utility of a player is simply given by
\begin{align*}
    \mathbb{E}_{\tau}[v(q)] &= \frac{p - q_0}{q_1 - q_0} [q_1 u^{11} + (1 - q_1) u^{01}] + \frac{q_1 - p}{q_1 - q_0} [q_0 u^{11} + (1 - q_0) u^{01}]\\
    &= \bigg(\frac{p - q_0}{q_1 - q_0} q_1 + \frac{q_1 - p}{q_1 - q_0} q_0\bigg) u^{11} + \bigg(\frac{p - q_0}{q_1 - q_0} (1 - q_1) + \frac{q_1 - p}{q_1 - q_0} (1 - q_0)\bigg) u^{01}\\
    &= p u^{11} + (1 - p) u^{01},
\end{align*}
which is the same as the expected utility of a player given the outcome $\tau$ with $supp(\tau) = \{p\}$, where no information is communicated to the receiver. This makes sense since the receiver would always take action $a = 1$ if $q_0 \geq \mu_R$, and thus we call all such outcomes the no-information outcome.
\begin{enumerate}
    \item Consider a 1-extremist. She would prefer to persuade the receiver to take action $a = 1$ as often as possible. Therefore she would prefer any outcome $\tau$ with $q_0 \geq \mu_R$, i.e., the no-information outcome, to all other outcomes\footnote{Her preferences over the outcomes $\tau$ with $q_0 < \mu_R$ do not matter since she prefers the no-information outcome to all such outcomes.}.
    \item Consider a 0-extremist. She would prefer to persuade the receiver to take action $a = 0$ as often as possible. The receiver would only take action $a = 0$ if his posterior belief is $q_0$ and $q_0 < \mu_R$. Since $\frac{\partial \tau(q_0)}{\partial q_0} > 0$ and $\frac{\partial \tau(q_0)}{\partial q_1} > 0$, she would prefer outcomes with higher $q_1$ and higher $q_0$ as long as $q_0 < \mu_R$. Clearly, her least preferred outcomes are those with $q_0 \geq \mu_R$, i.e., the no-information outcome.
    \item Consider a non-extremist. She would prefer any outcome $\tau$ with $q_0 \geq \mu_R$ to another outcome with $q_0 < \mu_R$ if and only if she prefers action $a = 1$ at $q_0$, as shown below:
    \begin{align*}
        p u^{11} + (1 - p) u^{01} &> \frac{p - q_0}{q_1 - q_0} [q_1 u^{11} + (1 - q_1) u^{01}] + \frac{q_1 - p}{q_1 - q_0} [q_0 u^{10} + (1 - q_0) u^{00}]\\
        q_0 (q_1 - p) (u^{11} - u^{10}) &> (1 - q_0) (q_1 - p) (u^{00} - u^{01})\\
        (1 - q_0) u^{01} + q_0 u^{11} &> (1 - q_0) u^{00} + q_0 u^{10}
    \end{align*}
    To see her preferences among outcomes $\tau$ with $q_0 < \mu_R$, note that
    \begin{align*}
        \frac{\partial \mathbb{E}_{\tau}[v(q)]}{\partial q_0} &= -\frac{q_1 - p}{(q_1 - q_0)^2}[q_1 u^{11} + (1 - q_1) u^{01} - q_0 u^{10} - (1 - q_0) u^{00}] + \frac{q_1 - p}{q_1 - q_0}(u^{10} - u^{00})\\
        &= -\frac{q_1 - p}{q_1 - q_0}(u^{00} - u^{01}) + \frac{q_1 - p}{(q_1 - q_0)^2}[q_1(u^{11} - u^{10}) - (1 - q_0)(u^{00} - u^{01})]\\
        &= -\frac{q_1 - p}{(q_1 - q_0)^2}[(1 - q_1) u^{01} + q_1 u^{11} - (1 - q_1) u^{00} - q_1 u^{10}]
    \end{align*}
    which is negative if and only if she prefers action $a = 1$ at $q_1$. Similarly, note that
    \begin{align*}
        \frac{\partial \mathbb{E}_{\tau}[v(q)]}{\partial q_1} &= -\frac{p - q_0}{(q_1 - q_0)^2}[q_1 u^{11} + (1 - q_1) u^{01} - q_0 u^{10} - (1 - q_0) u^{00}] + \frac{p - q_0}{q_1 - q_0}(u^{10} - u^{00})\\
        &= \frac{p - q_0}{q_1 - q_0}(u^{11} - u^{10}) - \frac{p - q_0}{(q_1 - q_0)^2}[q_1(u^{11} - u^{10}) - (1 - q_0)(u^{00} - u^{01})]\\
        &= \frac{p - q_0}{(q_1 - q_0)^2}[(1 - q_0) u^{00} + q_0 u^{10} - (1 - q_0) u^{01} - q_0 u^{11}]
    \end{align*}
   which is positive if and only if she prefers action $a = 0$ at $q_0$. Following these observations, we have:
    \begin{enumerate}
        \item If player $i$ is a conformist biased toward $a = 1$, she would prefer any outcome $\tau$ with $q_0 \geq \mu_R$ only to those with $\mu_i < q_0 < \mu_R$ since she prefers action $a = 1$ at $q_0$ only if $q_0 > \mu_i$\footnote{Note that if $\mu_i > \mu_R$, no outcome with $\mu_i < q_0 < \mu_R$ exists.}. Among outcomes $\tau$ with $q_0 \leq \mu_i$, she prefers ones with lower $q_0$ and higher $q_1$ since she prefers action $a = 1$ at $q_1$ and action $a = 0$ at $q_0$ if $q_0 < \mu_i$. 
        \item If player $i$ is a conformist biased toward $a = 0$, she would prefer any outcome $\tau$ with $q_0 < \mu_R$ to all those with $q_0 \geq \mu_R$ since she prefers action $a = 0$ at $q_0$. Among outcomes $\tau$ with $q_0 < \mu_R$, she prefers ones with higher $q_1$ since she prefers action $a = 0$ at $q_0$; also, she prefers ones with lower $q_0$ if $q_1 \geq \mu_i$ and higher $q_0$ if $q_1 < \mu_i$.
        \item If player $i$ is a contrarian biased toward $a = 1$, she would prefer any outcome $\tau$ with $q_0 \geq \mu_R$ to all outcomes with $q_0 < \mu_R$ since she prefers action $a = 1$ at $q_0$\footnote{Her preferences over the outcomes $\tau$ with $q_0 < \mu_R$ do not matter since she prefers the no-information outcome to all such outcomes.}.
        \item If player $i$ is a contrarian biased toward $a = 0$, she would prefer any outcome $\tau$ with $q_0 \geq \mu_R$ only to those with $q_0 < \min(\mu_i, \mu_R)$ since she prefers action $a = 1$ at $q_0$ only if $q_0 < \mu_i$. Among outcomes $\tau$ with $\mu_i \leq q_0 < \mu_R$, she prefers ones with higher $q_0$ and higher $q_1$ since she prefers action $a = 0$ at $q_1$ and action $a = 0$ at $q_0$ if $q_0 > \mu_i$\footnote{Note that if $\mu_i > \mu_R$, no outcome with $\mu_i < q_0 < \mu_R$ exists.}. 
    \end{enumerate}
\end{enumerate}
The above arguments imply that all we need to know about a player is her type, i.e., her indifference posterior belief $\mu_i$ and her preferred action at state $\omega = 1$.

\vspace{2mm}

\textbf{Proof of Proposition 4}:
Following 1, 3(c), and 3(d), if there exists a 1-extremist or a contrarian with $\mu_i > \mu_R$, the equilibrium outcome would be the no-information outcome since such a player would prefer the no-information outcome to all other outcomes. 

Following 3(a) and 3(d), if $A \neq \emptyset$ and there exists a contrarian with $\mu_{A^*} < \mu_i < \mu_R$, the equilibrium outcome would be the no-information outcome since $A^*$ would only prefer the outcomes with $q_0 \leq \mu_{A^*}$ to the no-information outcome while player $i$ would only prefer the outcomes with $\mu_i \leq q_0 < \mu_R$ to the no-information outcome. Note that these two sets of outcomes are disjoint.

Following 3(a) and 3(b), if all players are conformists, the equilibrium outcome would be the full-information outcome where $q_0 = 0$ and $q_1 = 1$ since this is the most-preferred outcome for all the players.

Let $D^*$ represent the player with the lowest bias among those in $D$ with $\mu_i < \mu_{A^*}$: $\mu_{D^*} = \max_{i \in D: \mu_i < \mu_{A^*}} \mu_i$. If $\{i \in D: \mu_i < \mu_{A^*}\} = \emptyset$, let $\mu_{D^*} = 0$.

Let $A^*_E$ represent the player with the highest bias among those in $A$ and closer to the receiver than $E^*$: $\mu_{A^*_E} = \min_{i \in A: i > E^*} \mu_i$. If $\{i \in A: i > E^*\} = \emptyset$, let $A^*_E = R$.

Following 2, 3(a), 3(b), and 3(d), if there exists no 1-extremists or contrarians with $\mu_i > \mu_{A^*}$, but there are players who are not conformists, that is, 0-extremists or contrarians with $\mu_i < \mu_{A^*}$, the equilibrium outcome would either be the no-information outcome or one with $\mu_{D^*} \leq q_0 \leq \mu_{A^*}$ and $q_1 = 1$.

\begin{itemize}
    \item If $A^* < E^*$, following 2 and 3(d), the best response of $E^*$ to any outcome induced by the preceding players with $\mu_{D^*} \leq q_0 \leq \mu_{A^*}$ and $q_1 = 1$ is to induce the outcome with $q_0 = \mu_{A^*_E}$ and $q_1 = 1$. However, following 3(a), $A^*$ would prefer the no-information outcome to such an outcome since by definition $\mu_{A^*} < \mu_{A^*_E}$. Thus, the only possible equilibrium outcome would be the no-information outcome.
    \item If $E^* < A^*$, following 2 and 3(d), the best response of any player in $D \cup E_0$ to any outcome induced by the preceding players with $\mu_{D^*} \leq q_0 \leq \mu_{A^*}$ and $q_1 = 1$ is to induce the outcome with $q_0 = \mu_{A^*}$ and $q_1 = 1$ since $A^* = A^*_E$. The best response of all conformists to any outcome induced by the preceding players with $\mu_{D^*} \leq q_0 \leq \mu_{A^*}$ and $q_1 = 1$ is to pass on that information. Moreover, all the players prefer any outcome with $\mu_{D^*} \leq q_0 \leq \mu_{A^*}$ to the no-information outcome. Thus, the equilibrium outcome would be the one with $q_0 = \mu_{A^*}$ and $q_1 = 1$.
\end{itemize} \hfill$\blacksquare$

\vspace{2mm}

\section{Appendix B: General Binary-Action Games}

In what follows, the sets $A$, $B$, $C$, $D$, $E_0$, and $E_1$ as well as players $A^*$ and $E^*$ are the same as those in section 5.2; that is, the definitions do not include player 1.

\textbf{Proof of Lemma 1}: I prove the result for the case that receiver is a conformist biased toward $a = 1$, i.e., $\alpha_R > 0$ and $0 < \omega_R < m$. The other cases can be proved similarly\footnote{By Assumption 2, if receiver is an absolute extremist, the equilibrium outcome of the reduced binary-state game corresponding to the original binary-action game is the full-information outcome.}.

As mentioned before, no-information outcome is the equilibrium outcome of the reduced binary-state game corresponding to the original binary-action game only if one of the following conditions holds:
\begin{enumerate}
    \item \textit{There exists a 1-extremist or a contrarian with $\omega_R < \omega_i < 1$ in the reduced game.} 
    
    Note that the only 1-extremists in the reduced game are the absolute 1-extremists in the original game. In the original game, any choice of $m_0$ and $m_1$ by player 1 does not change the type of absolute 1-extremists; it can only turn conformists or contrarians biased toward $a = 1$ into 1-extremists. This implies that contrarians with $m < \omega_i < 1$ either remain unchanged or are turned into 1-extremists. As a result, if there exists a 1-extremist or a contrarian with $m < \omega_i < 1$ in the reduced game, there would exist one in the subgame starting from player 2 in the original game as well, regardless of the choice of player 1, which leads to no-information outcome by Proposition 4.
    
    Any choice of $m_0$ and $m_1$ by player 1 that turns a contrarian with $\omega_R < \omega_i < m$ into an extremist satisfies $m_0 > \omega_R$; under the assumption that $\mathcal{S}_n = \mathcal{A}$, we must have $m_0 = m_1 = m$ which implies the no-information outcome. As a result, if there exists a contrarian with $\omega_R < \omega_i < m$ in the reduced game, either player 1 chooses the no-information outcome, or there would exist one in the subgame starting from player 2 in the original game as well, which leads to no-information outcome by Proposition 4. 
    \item \textit{$A \neq \emptyset$ and there exists a contrarian with $\omega_{A^*} < \omega_i < \omega_R$ in the reduced game.}
    
    Consider any choice of $m_0$ and $m_1$ by player 1 in the original game. If $m_0 \leq \omega_{A^*}$, we still have $A \neq \emptyset$ and contrarians with $\omega_{A^*} < \omega_i < \omega_R$ remain unchanged with $\omega_{A^*} < \omega_i$. If $m_0 > \omega_{A^*}$, all conformists in $A$ turn into 1-extremists; therefore, $A^*$ turns into a 1-extremist. As a result, if $A \neq \emptyset$ and there exists a contrarian with $\omega_{A^*} < \omega_i < m$ in the reduced game, either (i) $A \neq \emptyset$ and there would exist such a contrarian, or (ii) there would exist a 1-extremist in the subgame starting from player 2 in the original game, both of which lead to no-information outcome by Proposition 4.
    \item \textit{$D \cup E_0 \neq \emptyset$ and $A^* < E^*$ in the reduced game.}
    
    Note that the only 0-extremists in the reduced game are the absolute 0-extremists in the original game. In the original game, any choice of $m_0$ and $m_1$ by player 1 does not change the type of absolute 0-extremists; it can only turn conformists or contrarians biased toward $a = 0$ into 0-extremists. This implies that contrarians with $\omega_i < \omega_{A^*}$ either remain unchanged or are turned into 0-extremists. Therefore, for any choice of $m_0$ and $m_1$ by player 1, $E^*$ in the subgame starting from player 2 in the original game is the same as $E^*$ in the reduced game. 
    
    If $m_0 > \omega_{A^*}$, all conformists in $A$ turn into 1-extremists; therefore, $A^*$ turns into a 1-extremist. If $m_0 < \omega_{A^*}$, $A^*$ in the subgame starting from player 2 in the original game is the same as $A^*$ in the reduced game, and thus $A^* < E^*$. 
    
    As a result, if $A^* < E^*$ in the reduced game, either there would exist a 1-extremist, or $A^* < E^*$ in the subgame starting from player 2 in the original game, both of which lead to no-information outcome by Proposition 4. \hfill$\blacksquare$
\end{enumerate}

\vspace{2mm}

Bayes-plausibility implies that for any outcome of the game, or equivalently, any distribution of posterior means for the receiver $\tau$ with $supp(\tau) = \{m_0,m_1\}$, we have $\mathbb{E}_{\tau}[\omega] = m$. Without loss of generality, let $m_0 \leq m_1$ which implies $0 \leq m_0 \leq m \leq m_1 \leq 1$, $\tau(m_1) = \frac{m - m_0}{m_1 - m_0}$, and $\tau(m_0) = \frac{m_1 - m}{m_1 - m_0}$. Note that in the binary-action case, being smaller in the convex order boils down to being a mean preserving contraction. The set of binary-support distributions of posterior means which are mean preserving contractions of $\tau$ includes all $\tau'$ with $supp(\tau') = \{m'_0,m'_1\}$ such that $m_0 \leq m'_0 \leq m \leq m'_1 \leq m_1$ and $\mathbb{E}_{\tau'}[\omega] = m$. In other words, $\tau$ is more Blackwell-informative, the lower $m_0$ is and the higher $m_1$ is. 

\vspace{2mm}

\textbf{Proof of Lemma 4}: For the ease of exposition, suppose receiver takes $a = 1$ at $\omega = \omega_R$\footnote{This removes the necessity of Assumption 1.}. Given an outcome $\tau$ with $m_0 <\omega_R$, the expected utility of player 1 with $u_1(\omega,0) = 0$ and $u_1(\omega,1) = \alpha_1 \omega + \beta_1$ is given by
$$\mathbb{E}_{\tau}[u_1(\omega,a^*(\omega)] = \frac{m - m_0}{m_1 - m_0} (\alpha_1 m_1 + \beta_1)$$

Moreover, given any outcome $\tau$ with $m_0 \geq \omega_R$, the expected utility of a player 1 with $u_1(\omega,0) = 0$ and $u_1(\omega,1) = \alpha_1 \omega + \beta_1$ is simply given by
\begin{align*}
    \mathbb{E}_{\tau}[u_1(\omega,a^*(\omega)] &= \frac{m - m_0}{m_1 - m_0} [\alpha_1 m_1 + \beta_1] + \frac{m_1 - m}{m_1 - m_0} [\alpha_1 m_0 + \beta_1]\\
    &= \alpha_1\bigg(\frac{m - m_0}{m_1 - m_0} m_1 + \frac{m_1 - m}{m_1 - m_0} m_0\bigg) + \beta_1\\
    &= \alpha_1 m + \beta_1,
\end{align*}
which is the same as the expected utility of a player given the outcome $\tau$ with $supp(\tau) = \{m\}$, where no information is communicated to the receiver. This makes sense since the receiver would always take action $a = 1$ if $m_0 \geq \omega_R$, and thus we call all such outcomes the no-information outcome.

\begin{enumerate}
    \item Suppose player 1 is an absolute 1-extremist. She would prefer to persuade the receiver to take action $a = 1$ as often as possible. Therefore she would prefer any outcome $\tau$ with $m_0 \geq \omega_R$, i.e., the no-information outcome, to all other outcomes\footnote{Her preferences over the outcomes $\tau$ with $m_0 < \omega_R$ do not matter since she prefers the no-information outcome to all such outcomes.}.
    \item Suppose player 1 is an absolute 0-extremist. She would prefer to persuade the receiver to take action $a = 0$ as often as possible. The receiver would only take action $a = 0$ if his posterior mean is $m_0$ and $m_0 < \omega_R$. Since $\frac{\partial \tau(m_0)}{\partial m_0} > 0$ and $\frac{\partial \tau(m_0)}{\partial m_1} > 0$, she would prefer outcomes with higher $m_1$ and higher $m_0$ as long as $m_0 < \omega_R$. Clearly, her least preferred outcomes are those with $m_0 \geq \omega_R$, i.e., the no-information outcome.
    \item Suppose player 1 is a non-extremist. She would prefer any outcome $\tau$ with $m_0 \geq \omega_R$ to another outcome with $m_0 < \omega_R$ if and only if she prefers action $a = 1$ at $m_0$, as shown below:
    \begin{align*}
        \alpha_1 m + \beta_1 &> \frac{m - m_0}{m_1 - m_0} (\alpha_1 m_1 + \beta_1)\\
        \beta_1 m_1 - \alpha_1 m_0 m &> \beta_1 m - \alpha_1 m_1 m_0\\
        (m_1 - m) (\alpha_1 m_0 + \beta_1) &> 0
    \end{align*}
    To see her preferences among outcomes $\tau$ with $m_0 < \omega_R$, note that
    \begin{align*}
        \frac{\partial \mathbb{E}_{\tau}[u_1(\omega,a^*(\omega)]}{\partial m_0} &= -\frac{m_1 - m}{(m_1 - m_0)^2} (\alpha_1 m_1 + \beta_1)
    \end{align*}
    which is negative if and only if she prefers action $a = 1$ at $m_1$. Similarly, note that
    \begin{align*}
        \frac{\partial \mathbb{E}_{\tau}[u_1(\omega,a^*(\omega)]}{\partial m_1} &= -\frac{m - m_0}{(m_1 - m_0)^2} (\alpha_1 m_1 + \beta_1) + \frac{m - m_0}{m_1 - m_0} \alpha_1\\
        &= -\frac{m - m_0}{(m_1 - m_0)^2} (\alpha_1 m_0 + \beta_1)
    \end{align*}
   which is positive if and only if she prefers action $a = 0$ at $m_0$. Following these observations, we have:
    \begin{enumerate}
        \item If player 1 is a conformist biased toward $a = 1$, she would prefer any outcome $\tau$ with $m_0 \geq \omega_R$ only to those with $\omega_1 < m_0 < \omega_R$ since she prefers action $a = 1$ at $m_0$ only if $m_0 > \omega_1$\footnote{Note that if $\omega_1 > \omega_R$, no outcome with $\omega_1 < m_0 < \omega_R$ exists.}. Among outcomes $\tau$ with $m_0 \leq \omega_1$, she prefers ones with lower $m_0$ and higher $m_1$ since she prefers action $a = 1$ at $m_1$ and action $a = 0$ at $m_0$ if $m_0 < \omega_1$. 
        \item If player 1 is a conformist biased toward $a = 0$, she would prefer any outcome $\tau$ with $m_0 < \omega_R$ to all those with $m_0 \geq \omega_R$ since she prefers action $a = 0$ at $m_0$. Among outcomes $\tau$ with $m_0 < \omega_R$, she prefers ones with higher $m_1$ since she prefers action $a = 0$ at $m_0$; also, she prefers ones with lower $m_0$ if $m_1 \geq \omega_1$ and higher $m_0$ if $m_1 < \omega_1$.
        \item If player 1 is a contrarian biased toward $a = 1$, she would prefer any outcome $\tau$ with $m_0 \geq \omega_R$ to all outcomes with $m_0 < \omega_R$ since she prefers action $a = 1$ at $m_0$\footnote{Her preferences over the outcomes $\tau$ with $m_0 < \omega_R$ do not matter since she prefers the no-information outcome to all such outcomes.}.
        \item If player 1 is a contrarian biased toward $a = 0$, she would prefer any outcome $\tau$ with $m_0 \geq \omega_R$ only to those with $m_0 < \min(\omega_1, \omega_R)$ since she prefers action $a = 1$ at $m_0$ only if $m_0 < \omega_1$. Among outcomes $\tau$ with $\omega_1 \leq m_0 < \omega_R$, she prefers ones with higher $m_0$ and higher $m_1$ since she prefers action $a = 0$ at $m_1$ and action $a = 0$ at $m_0$ if $m_0 > \omega_1$\footnote{Note that if $\omega_1 > \omega_R$, no outcome with $\omega_1 < m_0 < \omega_R$ exists.}. 
    \end{enumerate}
\end{enumerate} \hfill$\blacksquare$

\vspace{2mm}

\textbf{Proof of Lemma 2}: I prove the result for the case that receiver is a conformist biased toward $a = 1$, i.e., $\alpha_R > 0$ and $0 < \omega_R < m$. The other cases can be proved similarly\footnote{By Assumption 2, if receiver is an absolute extremist, the equilibrium outcome of the binary-state game corresponding to the original binary-action game is the full-information outcome.}. 

No-information outcome is the equilibrium outcome of the binary-state game corresponding to the original binary-action game only if one of the following conditions holds:
\begin{enumerate}
    \item The equilibrium outcome of the reduced binary-state game corresponding to the original binary-action game is the no-information outcome. In this case, by Lemma 1, the equilibrium outcome of the original game is the no-information outcome as well.
    \item Player 1 is a 1-extremist or a contrarian with $\mu_R < \mu_1 < 1$ in the corresponding binary-state game. Following 1, 3(c), and 3(d) in the proof of Lemma 4, the equilibrium outcome of the original game would be the no-information outcome since player 1 would prefer the no-information outcome to all other outcomes. 
    \item $A \neq \emptyset$ and player 1 is a contrarian with $\omega_{A^*} < \omega_1 < \omega_R$ in the corresponding binary-state game. In the original game, following 3(d) in the proof of Lemma 4, player 1 would only prefer the outcomes with $\omega_1 \leq m_0 < \omega_R$ to the no-information outcome. However, any choice of $m_0$ and $m_1$ by player 1 with $m_0 \geq \omega_1$, turns $A^*$ into a 1-extremist in the subgame starting from player 2 in the original game, which leads to no-information outcome by Proposition 4. Therefore, the only possible equilibrium outcome in the original game would be the no-information outcome.  
    \item $D \cup E_0 \neq \emptyset$, $E^* < A^*$\footnote{If $A^* < E^*$, the equilibrium outcome of the reduced binary-state game corresponding to the original binary-action game is the no-information outcome, and we go back to condition 1.}, and $\omega_1 < \omega_{A^*}$ in the corresponding binary-state game. By Proposition 4, for any choice of $m_0$ and $m_1$ by player 1, the equilibrium outcome of the subgame starting from player 2 would be either the no-information outcome or $\tau$ with $supp(\tau) = \{\mu_{A^*} m_1 + (1-\mu_{A^*}) m_0, m_1\} = \{\omega_{A^*},m_1\}$. However, following 3(a) in the proof of Lemma 4, player 1 would prefer the no-information outcome since $\omega_1 < \omega_{A^*}$. Thus, the only possible equilibrium outcome in the original game would be the no-information outcome.
\end{enumerate} \hfill$\blacksquare$

\vspace{2mm}

\textbf{Proof of Lemma 3}: Since the no-information outcome is not the equilibrium outcome of the reduced binary-state game, not considering player 1, there exists no absolute 1-extremists or contrarians with $\omega_{A^*} < \omega_i$, and $A^*$ is closer to the receiver than all absolute 0-extremists and contrarians with $\omega_i < \omega_{A^*}$.

As mentioned before, given player 1's choice of distribution of posterior means $\tau_1$ with $supp(\tau_1) = \{m_0, m_1\}$, Proposition 4 characterizes the equilibrium outcome of the binary-state, binary-action subgame starting from player 2. Based on Proposition 4, the equilibrium of the subgame starting from player 2 is characterized by no-information outcome unless:
\begin{enumerate}
    \item all players are conformists: In this case, the equilibrium outcome $\tau$ of the binary-state, binary-action game starting from player 2 with $\widehat{\Omega} = \{m_0, m_1\}$ is characterized by $supp(\tau) = \{0,1\}$ (full-information outcome).
    
    All players in the binary-state, binary-action game starting from player 2 are conformists if and only if in the original game (i) all of them are conformists, and (ii) none of them turn into extremists given the choice of player 1, i.e., $m_0 \leq \omega_{A^*}$ and $m_1 \geq \omega_{B^*}$. If this is the case, $\tau$ implies that the the equilibrium outcome $\tau^*$ of the subgame starting from player 2 is characterized by $supp(\tau^*) = \{0\cdot m_1 + 1\cdot m_0, 1\cdot m_1 + 0\cdot m_0\} = \{m_0, m_1\}$.
    
    \item all players are not conformists but the non-conformist players are either 0-extremists or contrarians with $\omega_i < \omega_{A^*}$: In this case, the equilibrium outcome $\tau$ of the binary-state, binary-action game starting from player 2 with $\widehat{\Omega} = \{m_0, m_1\}$ is characterized by $supp(\tau) = \{\mu_{A^*},1\}$\footnote{As mentioned before, $\mu_{A^*} = \frac{m - m_0}{m_1 - m_0}$.}.
    
    All players in the binary-state, binary-action game starting from player 2 satisfy the above condition if and only if in the original game (i) all of them are conformists but at least one conformist biased toward $a = 0$ turns into a 0-extremist given the choice of player 1, i.e., $m_1 < \omega_{B^*}$, or there are players who are not conformists, (ii) none of the conformists biased toward $a = 1$ turn into 1-extremists given the choice of player 1, i.e., $m_0 \leq \omega_{A^*}$, and (iii) $A^*$ is closer to the receiver than all the new 0-extremists, i.e., conformists biased toward $a = 0$ with $\omega_i > m_1$. If this is the case, $\tau$ implies that the the equilibrium outcome $\tau^*$ of the subgame starting from player 2 is characterized by $supp(\tau^*) = \{\mu_{A^*}\cdot m_1 + (1-\mu_{A^*})\cdot m_0, 1\cdot m_1 + 0\cdot m_0\} = \{\mu_{A^*}, m_1\}$.
\end{enumerate} \hfill$\blacksquare$

\vspace{2mm}

\section{Appendix C: Proof of Proposition 3}

The first direction is straightforward. If $\tau \in \tilde{\Gamma}$, it implies that $\tau \in \Gamma_0$ and for every $i = 2, \ldots, n$, 
$$\sum_{\mu} \tau(\mu) v_i(\mu) \geq \sum_{\mu} \tau'(\mu) v_i(\mu),\forall \tau' \leq_{cx} \tau,$$
which, in turn, implies that for every $i = 2, \ldots, n$,
$$\sum_{\mu} \tau(\mu) v_i(\mu) \geq \sum_{\mu} \tau'(\mu) v_i(\mu),\ \forall \tau' \in \Gamma_{i+1}\ \text{s.t.}\ \tau' \leq_{cx} \tau^*,$$
where $\Gamma_{n+1} = \Gamma_0$. This can be used to show recursively that $\tau \in \Gamma_n, \tau \in \Gamma_{n-1}, \ldots, \tau \in \Gamma_2$. Therefore, $\tilde{\Gamma} \subseteq \Gamma_2$.

\vspace{2mm}

The following simple counterexample proves the other direction. However, it is based on the notation, concepts, and results introduced in section 5.1 and Appendix A. Consider a binary-state, binary-action game and let $n = 3$. Suppose player 2 is a 0-extremist while receiver and player 3 are conformists who are biased toward $a = 1$ with player 3 being higher biased, i.e, $0 < \mu_3 < \mu_R < p$. Note that the receiver takes action $a = 1$ at $\mu = \mu_R$ according to Assumption 1.

First of all, the set of Bayes plausible distributions of posteriors is given by

$$\Gamma_0 = \{\tau|supp(\tau) = \{q_0,q_1\}, 0 \leq q_0 \leq p \leq q_1 \leq 1, \mathbb{E}_{\tau}[q] = p\}.$$

As mentioned in Appendix B, in the binary-state binary-action case, being smaller in the convex order boils down to being a mean preserving contraction. The set of binary-support distributions of posteriors which are mean preserving contractions of $\tau$ includes all $\tau'$ with $supp(\tau') = \{q'_0,q'_1\}$ such that $q_0 \leq q'_0 \leq p \leq q'_1 \leq q_1$ and $\mathbb{E}_{\tau'}[q'] = p$. In other words, $\tau$ is more Blackwell-informative, the lower $q_0$ is and the higher $q_1$ is. 


Starting with player 3, and considering (\ref{IC_n}), we have
$$\Gamma_3 = \{\tau \in \Gamma_0|0 \leq q_0 \leq \mu_3\ \text{or}\ \mu_R \leq q_0 \leq p\}.$$
Now, continuing to player 2, and considering (\ref{IC_n-1}), we have
$$\Gamma_2 = \{\tau \in \Gamma_0|q_0 = \mu_3\ \text{or}\ \mu_R \leq q_0 \leq p\}.$$
However, 
$$\tilde{\Gamma} = \{\tau \in \Gamma_0|\mu_R \leq q_0 \leq p\}.$$ 
This is due to the fact that while player 2 prefers experiments with $q_0 = \mu_3$ to all less Blackwell-informative ones in $\Gamma_3$, i.e., those with $\mu_R \leq q_0 \leq p$, there exist less-Blackwell informative experiments in $\Gamma_0$, namely those with $\mu_3 < q_0 < \mu_R$, which she prefers even more. \hfill$\blacksquare$

\end{document}